\documentclass[prd,aps,nofootinbib,floatfix,11pt]{revtex4}
\usepackage{amsmath,graphicx,epsfig,amssymb,dsfont,mathtools,}
\usepackage[usenames]{color}
\usepackage{ulem} 
\usepackage{bigstrut}
\usepackage{slashed}
\usepackage{multirow}
\usepackage{subfigure}

\allowdisplaybreaks

\begin{document}\title{Light-Cone Sum Rules Analysis of $\Xi_{QQ^{\prime}}\to\Sigma_{Q^{\prime}}$ Weak Decays}

\author{Xiao-Hui Hu$^{1}$~\footnote{Email:huxiaohui@sjtu.edu.cn}, Yu-Ji Shi$^{2}$~\footnote{Corresponding author email:shiyuji92@126.com}}
\affiliation{$^{1}$ INPAC, SKLPPC, MOE KLPPC,
School of Physics and Astronomy, Shanghai Jiao Tong University, Shanghai 200240, China}
\affiliation{$^{2}$Helmholtz-Institut f\"ur Strahlen- und Kernphysik and Bethe Center for Theoretical Physics,\\ Universit\"at Bonn,
  53115 Bonn, Germany}

\begin{abstract}
As a continuation of our previous work, we investigate the weak decays of doubly-heavy baryons $\Xi_{QQ^{\prime}}$ into sextet $\Sigma_{Q^{\prime}}$ with light-cone sum rules. We calculate the form factors for these decays with the parallel light-cone distribution amplitudes of $\Sigma_{Q^{\prime}}$. Numerical results of these form factors are used to predict the decay widths and branching ratios of the corresponding semi-leptonic processes. Parametric uncertainties and theoretical analyses are also given in detail. We find that the decay widths of $\Xi_{cc}$ and $\Xi_{bc}$ decays are several orders of magnitude larger than those of $\Xi_{bb}$ decays. 
\end{abstract}
\maketitle

\section{Introduction}

The naive  quark model for mesons and baryons is very successful in explaining most hadrons observed on the experimental side.   Despite of these successes, there has been an important  renaissance in the hadron spectroscopy study in the past decades. On the one hand, a number of unexpected hadrons were observed most of which defy the standard quark-anti-quark interpretation for mesonic states, and three-quark scenario for baryonic states. These observations of hadron exotics have triggered tremendous theoretical studies, but no conclusive results are obtained yet.   On the other side, not all predicted particles by the quark model are well established on the experimental side. Among the latter category,  doubly-charmed baryon is an intriguing example that has been discussed in the recent years.   In 2017, the LHCb collaboration announced the observation of the ground state doubly-charmed baryon $\Xi_{cc}^{++}$ whose mass is given as~\cite{Aaij:2017ueg}
\begin{equation}
m_{\Xi_{cc}^{++}}=(3621.40\pm0.72\pm0.27\pm0.14)\ {\rm MeV}.\label{eq:LHCb_measurement}
\end{equation}
This newly observed particle was reconstructed from the decay channel $\Lambda_{c}^{+}K^{-}\pi^{+}\pi^{+}$,  which had been predicted in Ref.~\cite{Yu:2017zst}. One year later LHCb announced their measurement on $\Xi_{cc}^{++}$ lifetime~\cite{Aaij:2018wzf} as well as the confirmation of  $\Xi_{cc}^{++}$ in the   $\Xi_{c}^{+}\pi^{+}$ final state~\cite{Aaij:2018gfl}. Undoubtedly, experimentalists will continue  to search for other heavier particles included in the doubly-heavy baryon spectroscopy~\cite{Traill:2017zbs,Cerri:2018ypt,Aaij:2019jfq}. {Tab.~\ref{tab:JPC} lists various doubly heavy baryons as well as their corresponding quantum numbers. On the other hand, the great progress  in  experiments also arouses theoretical analyses, together with which a deeper understanding of hadron spectrum might be achieved in future. 
To date there have been some  theoretical studies which aim to understand the dynamic and spectroscopy properties of the doubly-heavy baryon states~\cite{Wang:2017mqp,Meng:2017udf,Wang:2017azm,
Gutsche:2017hux,Li:2017pxa,Guo:2017vcf,Xiao:2017udy,Sharma:2017txj,Ma:2017nik,Hu:2017dzi,Shi:2017dto,Yao:2018zze,Yao:2018ifh,
Ozdem:2018uue,Ali:2018ifm,Dias:2018qhp,Zhao:2018mrg,Xing:2018bqt,
Ali:2018xfq,Liu:2018euh,Xing:2018lre,Bediaga:2018lhg,Wang:2017vnc,Dhir:2018twm,Berezhnoy:2018bde,Jiang:2018oak,Zhang:2018llc,Li:2018bkh,Gutsche:2018msz}, however, a comprehensive description of the decay mechanism of doubly heavy baryons is not established yet. 
\begin{table*}[!htb]
\footnotesize
\caption{Quantum numbers and quark content for the ground state of doubly heavy baryons.  The $S_{h}^{\pi}$ denotes the spin of the heavy quark system. The light quark $q$ corresponds to $u,d$ quark.   }\label{tab:JPC}
\begin{center}
\begin{tabular}{cccc|cccccc} \hline \hline
Baryon      & Quark Content  &  $S_h^\pi$  &$J^P$   & Baryon & Quark Content &   $S_h^\pi$  &$J^P$   \\ \hline  
$\Xi_{cc}$ & $\{cc\}q$  & $1^+$ & $1/2^+$ &   $\Xi_{bb}$ & $\{bb\}q$  & $1^+$ & $1/2^+$ & \\  
$\Xi_{cc}^*$ & $\{cc\}q$  & $1^+$ & $3/2^+$ &   $\Xi_{bb}^*$ & $\{bb\}q$  & $1^+$ & $3/2^+$ & \\ \hline  
$\Omega_{cc}$ & $\{cc\}s$  & $1^+$ & $1/2^+$ &   $\Omega_{bb}$ & $\{bb\}s$  & $1^+$ & $1/2^+$ & \\  
$\Omega_{cc}^*$ & $\{cc\}s$  & $1^+$ & $3/2^+$ &   $\Omega_{bb}^*$ & $\{bb\}s$  & $1^+$ & $3/2^+$ &  \\ \hline  
$\Xi_{bc}'$ & $\{bc\}q$  & $0^+$ & $1/2^+$ &   $\Omega_{bc}'$ & $\{bc\}s$  & $0^+$ & $1/2^+$ & \\  
$\Xi_{bc}$ & $\{bc\}q$  & $1^+$ & $1/2^+$ &   $\Omega_{bc}$ & $\{bc\}s$  & $1^+$ & $1/2^+$ & \\  
$\Xi_{bc}^*$ & $\{bc\}q$  & $1^+$ & $3/2^+$ &   $\Omega_{bc}^*$ & $\{bc\}s$  & $1^+$ & $3/2^+$ & 
 \\ \hline \hline
\end{tabular}
\end{center}
\end{table*}

One of the ideal platforms for studying doubly heavy baryon is its semi-leptonic decay, where a doubly heavy baryon transforms to a singly heavy baryon through a weak current. All the QCD dynamics in such processes are packed in a single transition matrix element which is suitable for a pure theoretical study. 
On the other hand, unlike the singly heavy baryon decays such as $\Lambda_b\to p$, the doubly to singly heavy baryon transitions contain a heavy quark in the final state, which means that the final state is most likely a low energy state. So the heavy quark symmetry, as a static property of heavy hadrons, may still emerge in such processes. Thus the transition matrix element is expected to be dominated by the dynamics describing how the light degrees of freedom response to the change of the heavy quark velocity. 
In our previous work~\cite{Shi:2019fph},  we have given this estimation according to our phenomenological study on the semi-leptonic weak decays of $\Xi_{QQ^{\prime}}\to \Lambda_{Q^{\prime}}$ within the framework of light cone sum rules (LCSR).  In that work, the final states $\Lambda_{Q^{\prime}}$ belong to SU(3) antitriplets $\bar{3}$. According to the flavor SU(3) symmetry,  the singly heavy baryons are classified by irreducible SU(3) representations $3\otimes3=6\oplus\bar{3}$. Thus besides antitriplets, sextet baryons including $\Sigma_{Q}$, $\Xi_{Q}^{\prime}$ and $\Omega_{Q}$ can also be final states of $\Xi_{QQ^{\prime}}$ decays. In this paper we analyze the transition $\Xi_{QQ^{\prime}}\to\Sigma_{Q^{\prime}}$ by LCSR, where the $u,d$ quark mass of the final state can be neglected so that the corresponding LCDAs are well defined. 
It can be expected that this work will provide a powerful guide for future experiments on doubly heavy baryon decays.

The transition matrix element of $\Xi_{QQ^{\prime}}\to\Sigma_{Q^{\prime}}$ is parametrized by six form factors. In previous literatures \cite{Wang:2017mqp,Zhao:2018mrg,Shi:2019hbf}, these form factors were calculated by light-front quark model (LFQM) or QCD sum rules (QCDSR). By LFQM, the two spectator quarks are treated as a diquark, which can simplify the calculation of the transition matrix element~\cite{Zhao:2018mrg,Xing:2018lre}. However, {\color{red}{since}} the dynamics in the diquark system are smeared, the diquark approximation may introduce some systematic uncertainties. In the QCDSR approach, to evaluate the transition matrix element one needs to introduce a three-point correlation function. This correlation function is calculated by operator product expansion (OPE), while the truncated OPE may lead to the potential irregularities. In addition, such calculation is complex especially when high dimensional operator condensations are included. Most of them must heavily depend on numerical evaluations, where large systematic uncertainties are also inevitable \cite{Colangelo:2000dp}.  In this work, to avoid the above problems, we \textcolor{red}{apply} LCSR to analyze doubly-heavy baryon transition form factors. In the framework of LCSR, the non-perturbative dynamics of the quarks and gluons in the baryons are described by the light-cone distribution amplitudes(LCDAs). While only a two-point correlation function is needed for calculating these form factors. We hope our study basing on LCSR can help people to understand the heavy baryon LCDAs, and examine their justifiability.

This paper is arranged as follows. In Sec.\ref{sec:lc_sum_rules}, we parametrize the matrix elements of transitions $\Xi_{QQ^{\prime}}\to\Sigma_{Q^{\prime}}$ with form factors $f_{i}$ and $g_{i}$. Then we introduce the LCSR approach for calculating these form factors. The numerical results are presented in Sec.\ref{sec:numerical results}, including the results of form factors, the predictions of decay widths and branching ratios of doubly heavy baryon semi-leptonic decays.  Sec.\ref{sec:conclusions} is a brief summary of this work.

\section{Transition Form Factors in light-cone sum rules}
\label{sec:lc_sum_rules}

\subsection{Form Factors}

The transition $\Xi_{QQ^{\prime}}\to\Sigma_{Q^{\prime}}$ matrix elements induced by $(V-A)^{\mu}$ current are expressed in terms of six form factors, three for vector current while the other three for axial-vector current
\begin{eqnarray}
&&\langle{\Sigma_{Q^{\prime}}}(p_{\Sigma},s_{\Sigma})|(V-A)^{\mu}|{\Xi_{QQ^{\prime}}}(p_{\Xi},s_{\Xi})\rangle \nonumber \\ & = & \bar{u}_{\Sigma}(p_{\Sigma},s_{\Sigma})\bigg[\gamma^{\mu}f_{1}(q^{2})+i\sigma^{\mu\nu}\frac{q_{\nu}}{m_{\Xi}}f_{2}(q^{2})+\frac{q^{\mu}}{m_{\Xi}}f_{3}(q^{2})\bigg]u_{\Xi}(p_{\Xi},s_{\Xi})\nonumber \\
&  &- \bar{u}_{\Sigma}(p_{\Sigma},s_{\Sigma})\bigg[\gamma^{\mu}g_{1}(q^{2})+i\sigma^{\mu\nu}\frac{q_{\nu}}{m_{\Xi}}g_{2}(q^{2})+\frac{q^{\mu}}{m_{\Xi}}g_{3}(q^{2})\bigg]\gamma_{5}u_{\Xi}(p_{\Xi},s_{\Xi}), 
\label{eq:parameterization1}
\end{eqnarray}
where $q^{\mu}=p_{\Xi}^{\mu}-p_{\Sigma}^{\mu}$ is transferring momentum. 

To simplify the extraction of form factors, another parameterizing scheme can be applied to the same matrix element, which is more applicable in the frame work of LCSR.
\begin{eqnarray}
&&\langle{\Sigma_{Q^{\prime}}}(p_{\Sigma},s_{\Sigma})|(V-A)^{\mu}|{\Xi_{QQ^{\prime}}}(p_{\Xi},s_{\Xi})\rangle \nonumber \\ & = & \bar{u}_{\Sigma}(p_{\Sigma},s_{\Sigma})\bigg[F_{1}(q^{2})\gamma^{\mu}+F_{2}(q^{2})p_{\Sigma}^{\mu}+F_{3}(q^{2})p_{\Xi}^{\mu}\bigg]u_{\Xi}(p_{\Xi},s_{\Xi})\nonumber \\
&  &- \bar{u}_{\Sigma}(p_{\Sigma},s_{\Sigma})\bigg[G_{1}(q^{2})\gamma^{\mu}+G_{2}(q^{2})p_{\Sigma}^{\mu}+G_{3}(q^{2})p_{\Xi}^{\mu}\bigg]\gamma_{5}u_{\Xi}(p_{\Xi},s_{\Xi}). 
\label{eq:parameterization2}
\end{eqnarray}
The form factors $F_i$ and $G_i$ in the above equation have the following relations with the $f_i$ and $g_i$ defined in Eq. (\ref{eq:parameterization1})
\begin{eqnarray}
f_1(q^2) &=& F_1(q^2)+\frac{1}{2}(m_{\Xi}+m_{\Sigma})(F_2(q^2)+F_3(q^2)), \nonumber \\
f_2(q^2) &=& \frac{1}{2}m_{\Xi}(F_2(q^2)+F_3(q^2)), \nonumber \\
f_3(q^2) &=& \frac{1}{2}m_{\Xi}(F_3(q^2)-F_2(q^2)), \nonumber \\
g_1(q^2) &=& G_1(q^2)-\frac{1}{2}(m_{\Xi}-m_{\Sigma})(G_2(q^2)+G_3(q^2)), \nonumber \\
g_2(q^2) &=& \frac{1}{2}m_{\Xi}(G_2(q^2)+G_3(q^2)), \nonumber \\
g_3(q^2) &=& \frac{1}{2}m_{\Xi}(G_3(q^2)-G_2(q^2)).
\end{eqnarray}

\subsection{Light-Cone Distribution Amplitudes of $\Sigma_{Q}$}

The light-cone distribution functions of sextet baryons with spin-parities $J^{P}=1/2^{+}$ were provided in Ref.~\cite{Ali:2012pn}, where the LCDAs are calculated by QCDSR in the heavy quark mass limit. In Ref.~\cite{Ali:2012pn}, the sextet LCDAs are classified by the total spin polarization of the two light quarks. If the polarization vector is parallel to the light-cone plane, four parallel LCDAs are defined by four parallel currents respectively
\begin{eqnarray}
\frac{\bar{v}^{\mu}}{v_{+}}\langle0|[q_{1}^{T}(t_{1})C\slashed nq_{2}(t_{2})]Q_{\gamma}(0)|\Sigma_{Q}(v)\rangle & =\frac{1}{\sqrt{3}}\psi^{n}_{\parallel}(t_{1},t_{2})f^{(1)}\epsilon_{\parallel}^{\mu} u_{\gamma},\nonumber\\
\frac{i\bar{v}^{\mu}}{2}\langle0|[q_{1}^{T}(t_{1})C\sigma\textcolor{red}{_{\alpha\beta}}q_{2}(t_{2})]Q_{\gamma}(0)\bar{n}^{
\textcolor{red}{\alpha}}n^{\textcolor{red}{\beta}}|\Sigma_{Q}(v)\rangle & =\frac{1}{\sqrt{3}}\psi^{n\bar{n}}_{\parallel}(t_{1},t_{2})f^{(2)}\epsilon_{\parallel}^{\mu}u_{\gamma},\nonumber\\
\bar{v}^{\mu}\langle0|[q_{1}^{T}(t_{1})Cq_{2}(t_{2})]Q_{\gamma}(0)|\Sigma_{Q}(v)\rangle & =\frac{1}{\sqrt{3}}\psi^{1}_{\parallel}(t_{1},t_{2})f^{(2)}\epsilon_{\parallel}^{\mu}u_{\gamma},\nonumber\\
-v_{+}\bar{v}^{\mu}\langle0|[q_{1}^{T}(t_{1})C\bar{\slashed n}q_{2}(t_{2})]Q_{\gamma}(0)|\Sigma_{Q}(v)\rangle & =\frac{1}{\sqrt{3}}\psi^{\bar{n}}_{\parallel}(t_{1},t_{2})f^{(1)}\epsilon_{\parallel}^{\mu}u_{\gamma}.\label{LCDAdef}
\end{eqnarray}
Here $\gamma$  is a Dirac spinor index. $n$ and $\bar n$ are the two light-cone vectors, while ${\bar v}^{\mu}=\frac{1}{2}(\frac{n^{\mu}}{v_{+}}-v_{+}\bar{n}^{\mu})$. $t_{i}$ are the distances between the $i$th light quark and the origin along the direction of $n$. The space-time coordinates of the light quarks can be expressed as $t_{i}n^{\mu}$. The four-velocity of $\Sigma_{Q}$ is defined by light-cone coordinates $v^{\mu}=\frac{1}{2}(\frac{n^{\mu}}{v_{+}}+v_{+}\bar{n}^{\mu})$, where $v_+=1$ since we choose the rest frame of $\Sigma_{Q}$ in this work.  When $q_1$ and $q_2$ are the same, an extra factor $\sqrt{2}$ should times on the right hand side of Eq.~(\ref{LCDAdef}). 
Note that $Q$ should be effective heavy quark field satisfying $\slashed v Q=Q$.  However, at the leading order we will not distinguish it from the original field. $\{\psi^{n}_{\parallel}, \psi^{n\bar{n}}_{\parallel},\psi^{1}_{\parallel}, \psi^{\bar{n}}_{\parallel}\}$ denote the four parallel LCDAs with different twists, which correspond to $\{\psi_{2}, \psi^{\sigma}_{3},\psi^{s}_{3}, \psi_{4}\}$. 

As shown in Ref.~\cite{Ali:2012pn}, if the polarization vector is transversal to the light-cone plane, one needs to introduce other four transversal LCDAs by four transversal currents. However, due to the fact that the form factors are only functions of $q^2$, one can choose any polarization configuration of the sextet baryon to extract the form factors. In fact, as argued by Ref.~\cite{Ali:2012pn}, $\psi_{\parallel}$ and $\psi_{\perp}$ are not independent with each other, they are related by the parameter $A$ which will be shown in Tab.~\ref{modelparam}. When $A=1/2$, they are the same. Thus in this work, we only use the parallel LCDAs defined in Eq.~(\ref{LCDAdef}) by choosing the light part polarization vector of the sextet baryon being parallel to the light-cone plane.
With the four parallel LCDAs, the matrix element $\epsilon_{abc}\langle\Sigma_{Q}(v)|\bar{q}_{1k}^{a}(t_{1})\bar{q}_{2i}^{b}(t_{2})\bar{Q}_{\gamma}^{c}(0)|0\rangle$ is expressed as
\begin{eqnarray}
\epsilon_{abc}\langle\Sigma_{Q}(v)|\bar{q}_{1k}^{a}(t_{1})\bar{q}_{2i}^{b}(t_{2})\bar{Q}_{\gamma}^{c}(0)|0\rangle & =&\frac{1}{8}v_{+}\psi_{\parallel}^{n*}(t_{1},t_{2})f^{(1)}(\bar{u}_{\Sigma}\bar{\slashed v}\gamma_{5})_{\gamma}(C^{-1}\bar{\slashed n})_{ki}\nonumber\\
 & &-\frac{1}{8}\psi_{\parallel}^{n\bar{n}*}(t_{1},t_{2})f^{(2)}(\bar{u}_{\Sigma}\bar{\slashed v}\gamma_{5})_{\gamma}(C^{-1}i\sigma\textcolor{red}{^{\alpha\beta}})_{ki}\bar{n}\textcolor{red}{_{\alpha}}n\textcolor{red}{_{\beta}}\nonumber\\
 & &+\frac{1}{4}\psi_{\parallel}^{1*}(t_{1},t_{2})f^{(2)}(\bar{u}_{\Sigma}\bar{\slashed v}\gamma_{5})_{\gamma}(C^{-1})_{ki}\nonumber\\
 & &-\frac{1}{8v_{+}}\psi_{\parallel}^{\bar{n}*}(t_{1},t_{2})f^{(1)}(\bar{u}_{\Sigma}\bar{\slashed v}\gamma_{5})_{\gamma}(C^{-1}\slashed n)_{kl},\label{quarksmatrix}
\end{eqnarray}
where the color indexes $a,\ b,\ c$ are explicitly summed over. The Fourier transformation of the LCDAs are
\begin{eqnarray}
\psi(x_{1},x_{2})=\int_{0}^{\infty}d\omega_{1}d\omega_{2}e^{-i\omega_{1}t_{1}}e^{-i\omega_{2}t_{2}}\psi(\omega_{1},\omega_{2}),
\end{eqnarray}
where $\omega_{1}$ and $\omega_{2}$ are the momentum of the two light
quarks, which are along the light-cone direction. The total diquark momentum has the same direction with magnitude
$\omega=\omega_{1}+\omega_{2}$. Note that $x_{1}=t_{1}n$ , $x_{2}=t_{2}n$
\begin{eqnarray}
\psi(t_{1},t_{2}) &=&\int_{0}^{\infty}d\omega d\omega_{2}e^{-i\omega t_{1}}e^{-i\omega_{2}(t_{2}-t_{1})}\psi(\omega_{1},\omega_{2}), \\
\psi(0,t_{2})&=&\int_{0}^{\infty}d\omega\omega\int_{0}^{1}due^{-i\bar{u}\omega v\cdot x_{2}}\psi(\omega,u),
\end{eqnarray}
where $\omega_2=(1-u)\omega=\bar u \omega$. $t_{i}$ are expressed in Lorentz invariant form $t_{i}=v\cdot x_{i}$.
Although Ref.~\cite{Ali:2012pn} only provides LCDAs of bottom baryons, according to the argument given in Ref.~\cite{Shi:2019fph}, one can safely apply these LCDAs for charm baryons in heavy quark limit. In this work both $\Sigma_{b}$ and $\Sigma_{c}$ are described by the same LCDAs given in Ref.~\cite{Ali:2012pn}, which are expressed as
\begin{eqnarray}
 \psi_{2}(\omega,u) &=& \omega^2 \bar u u\sum_{n=0}^{2}
 \frac{a_{n}}{\epsilon_{n}^{4}}\frac{C_{n}^{3/2}(2u-1)}{|C_{n}^{3/2}|^2}e^{-\omega/\epsilon_{n}}\,,
\nonumber\\
 \psi_{4}(\omega,u) &=& \sum_{n=0}^{2}
 \frac{a_{n}}{\epsilon_{n}^{2}}\frac{C_{n}^{1/2}(2u-1)}{|C_{n}^{1/2}|^2}e^{-\omega/\epsilon_{n}}\,,
\nonumber\\
 \psi_{3}^{\sigma,s}(\omega,u) &=& \frac{\omega}{2}\sum_{n=0}^{2}
 \frac{a_{n}}{\epsilon_{n}^{3}}\frac{C_{n}^{1/2}(2u-1)}{|C_{n}^{1/2}|^2}e^{-\omega/\epsilon_{n}}\,,
\label{SR:pert}
\end{eqnarray}
with 
\begin{eqnarray}
   |C_{n}^{\lambda}|^2 &=& \int_0^{1}[C_{n}^{\lambda}(2u-1)]^2\, \,,
\end{eqnarray}
where $C_{0}^{\lambda}(x)=1$, $C_{1}^{\lambda}(x)=2\lambda x$ and $C_{2}^{\lambda}(x)=2\lambda(1+\lambda) x^2-\lambda$.
The parameters in Eq.~\eqref{SR:pert} are collected 
in Tab.~\ref{modelparam}. The parameter $A$ is chosen to be around $1/2$ and also make sure that $\epsilon_i$\textcolor{red}{s} are non-negative. In this work, we simply choose it to be $A=1/2$.
\begin{table}
\caption{
Parameters for the parallel LCDAs of $\Sigma_{b}$ in 
Eqs.~(\ref{SR:pert}). A replacement $A \to 1-A$ is made for transversal LCDAs \cite{Ali:2012pn}.
2, $3\sigma$, 3s and 4 are twist notations. 
}   
\label{modelparam}
\begin{center}
\def\arraystretch{1.5}
\begin{tabular}{c|c|p{2cm}p{2cm} p{2cm} p{2cm} p{2cm} p{2cm}}
\hline\hline \multirow{6}{*}{\mbox{\Large $ \Sigma_b $}}    
&twist& $a_0$ & $a_1$ & $a_2$ & $\varepsilon_0$[GeV] & $\varepsilon_1$[GeV] & $\varepsilon_2$[GeV]  \\ 
\cline{2-8} 
&$2      $  & $  1 $ & $ -  $ & $ \frac{6.4 A}{A+0.44}  $& $  \frac{1.4 A+0.6}{A+5.7} $& $ -  $ &$ \frac{0.32 A}{A-0.17}  $\\
&$3 s    $  & $  1 $ & $ -  $ & $ \frac{0.12 A-0.08}{A-1.4}  $& $ \frac{0.56 A-0.77}{A-2.6}  $& $ -  $ &$ \frac{0.25 A-0.16}{A+0.41}  $\\
&$3\sigma     $  & $ -  $ & $ 1  $ & $ -  $& $ -  $& $ \frac{  0.35 A -0.43   }{A-1.2 }  $ &$ -  $\\
&$4      $  & $ 1  $ & $ -  $ & $ \frac{  -0.07 A - 0.05   }{A +0.34 }  $& $ \frac{  0.65 A+0.22   }{A+1 }  $& $ -  $ &$ \frac{  5.5 A+3.8   }{A +29 }  $\\
 \hline\hline 
\end{tabular}
\end{center}
\end{table}

\subsection{Light-Cone Sum Rules Framework}	

According to the framework of LCSR, to deal with the transition defined in Eq. (\ref{eq:parameterization2}), one needs to construct a two point correlation function
\begin{equation}
\Pi_{\mu}(p_{\Sigma},q) =i\int d^{4}xe^{iq\cdot x}\langle\Sigma_{Q^{\prime}}(p_{\Sigma})|T\{J^{V-A}_{\mu}(x)\bar{J}_{\Xi_{QQ^{\prime}}}(0)\}|0\rangle\label{eq:corrfunc}.
\end{equation}
The two currents $J^{V-A},\ J_{\Xi_{QQ^{\prime}}}$ are $V-A$ current and the $\Xi_{QQ^{\prime}}$ interpolating current respectively
\begin{equation}
J^{V-A}_{\mu}(x) =\bar{q}_{e}\gamma_{\mu}(1-\gamma_{5})Q_{e},
\end{equation}
while for $Q=Q^{\prime}=b,\ c$
\begin{equation}
J_{\Xi_{QQ}} =\epsilon_{abc}(Q_{a}^{T}C\gamma\textcolor{red}{^{\nu}}Q_{b})\gamma\textcolor{red}{_{\nu}}\gamma_{5}q^{\prime}_{c},
\end{equation}
for $Q=b,\ Q^{\prime}=c$
\begin{equation}
J_{\Xi_{bc}} =\frac{1}{\sqrt{2}}\epsilon_{abc}(b_{a}^{T}C\gamma\textcolor{red}{^{\nu}}c_{b}
+c_{a}^{T}C\gamma\textcolor{red}{^{\nu}}b_{b})\gamma\textcolor{red}{_{\nu}}\gamma_{5}q^{\prime}_{c}.
\end{equation}

In the frame work of LCSR, the correlation function Eq. (\ref{eq:corrfunc}) is calculated both at hadron level and QCD level. Then the results at the two levels can be matched according to quark-hadron duality. At hadron level, by inserting a complete set of baryon states including both positive and negative parity states between $J^{V-A}$ and $J_{\Xi_{QQ^{\prime}}}$, and using the definition of $\Xi_{QQ^{\prime}}^{P+}$ and $\Xi_{QQ^{\prime}}^{P-}$ decay constants  $f^{+}_{\Xi}$ and $f^{-}_{\Xi}$ 
\begin{eqnarray}
\langle {\Xi^{P+}_{QQ^{\prime}}}(p_{ \Xi},s)|\bar{J}_{\Xi_{QQ^{\prime}}}(0)|0\rangle&=&f^{+}_{\Xi}\bar{u}_{\Xi}(p_{\Xi},s) \nonumber\\
\langle {\Xi^{P-}_{QQ^{\prime}}}(p_{ \Xi},s)|\bar{J}_{\Xi_{QQ^{\prime}}}(0)|0\rangle&=&-i{\gamma_5}f^{-}_{\Xi}\bar{u}_{\Xi}(p_{\Xi},s),
\end{eqnarray}
After inserting the positive and negative parity hadron states respectively, the correlation function induced by the vector current $\bar q \gamma^\mu Q$ has the following forms
\begin{eqnarray}
\Pi_{\mu,V}^{hadron}(p_{\Sigma},q)^{+} & =&-\frac{f^{+}_{\Xi}}{(q+p_{\Sigma})^{2}-m_{\Xi}^{2}}\bar{u}_{\Sigma}(p_{\Sigma})\nonumber\\
&&\times[F_{1}^{+}(q^{2})\gamma_{\mu}+F_{2}^{\textcolor{red}{+}}(q^{2})p_{\Sigma\mu}+F_{3}^{+}(q^{2})p_{\Xi\mu}](\slashed q+\slashed p_{\Sigma}+m_{\Xi})+\cdots \nonumber\\
&=&-\frac{f^{+}_{\Xi}}{(q+p_{\Sigma})^{2}-m_{\Xi}^{2}}\bar{u}_{\Sigma}(p_{\Sigma})\Big[F_{1}^{+}(q^{2})(m_{\Xi}-m_{\Sigma})\gamma_{\mu}+(m_{\Xi}+m_{\Sigma})F_{3}^{+}(q^{2})q_{\mu}\nonumber\\
&&+[(m_{\Sigma}^{2}+m_{\Xi}m_{\Sigma})(F_{2}^{+}(q^{2})+F_{3}^{+}(q^{2}))+2m_{\Sigma}F_{1}^{+}(q^{2})]v_{\mu} \nonumber\\
 & &+F_{1}^{+}(q^{2})\gamma_{\mu}\slashed q+m_{\Sigma}(F_{2}^{+}(q^{2})+F_{3}^{+}(q^{2}))v_{\mu}\slashed q+F_{3}^{+}(q^{2})q_{\mu}\slashed q\Big]
 + \cdots,\label{FormextractP}
\end{eqnarray}
\begin{eqnarray}
\Pi_{\mu,V}^{hadron}(p_{\Sigma},q)^{-} 
&=&-\frac{f^{-}_{\Xi}}{(q+p_{\Sigma})^{2}-m_{\Xi}^{2}}\bar{u}_{\Sigma}(p_{\Sigma})\nonumber\\
&&\times[F^{-}_{1}(q^{2})\gamma_{\mu}+F^{-}_{2}(q^{2})p_{\Sigma\mu}+F^{-}_{3}(q^{2})p_{\Xi\mu}](-\slashed q-\slashed p_{\Sigma}+m_{\Xi})+\cdots\nonumber\\
&=&-\frac{f^{-}_{\Xi}}{(q+p_{\Sigma})^{2}-m_{\Xi}^{2}}\bar{u}_{\Sigma}(p_{\Sigma})\Big[F_{1}^{-}(q^{2})(m_{\Xi}+m_{\Sigma})\gamma_{\mu}+(m_{\Xi}-m_{\Sigma})F_{3}^{-}(q^{2})q_{\mu}\nonumber\\
&&+[(-m_{\Sigma}^{2}+m_{\Xi}m_{\Sigma})(F_{2}^{-}(q^{2})+F_{3}^{-}(q^{2}))-2m_{\Sigma}F_{1}{}^{-}(q^{2})]v_{\mu} \nonumber\\
 & &-F_{1}^{-}(q^{2})\gamma_{\mu}\slashed q-m_{\Sigma}(F_{2}^{-}(q^{2})+F_{3}^{-}(q^{2}))v_{\mu}\slashed q-F_{3}^{-}(q^{2})q_{\mu}\slashed q\Big]+ \cdots,\label{FormextractN}
\end{eqnarray}
where the ellipses stand for the contribution from continuum spectra $\rho^{h}$ above the threshold $s_{th}$, which has the integral form 
\begin{eqnarray}
\int_{s_{th}}^{\infty}ds\frac{\rho^{h}(s,q^{2})}{s-p_{\Xi}^{2}}.\label{correHadron}
\end{eqnarray}
The total hadron level correlation function is contributed both from Eqs. (\ref{FormextractP}) and (\ref{FormextractN})
\begin{eqnarray}
\Pi_{\mu,V}^{hadron}(p_{\Sigma},q)=\Pi_{\mu,V}^{hadron}(p_{\Sigma},q)^{+}+\Pi_{\mu,V}^{hadron}(p_{\Sigma},q)^{-}.\label{Formextract}
\end{eqnarray}
The correlation function induced by the axial-vector current $\bar q \gamma^\mu\gamma^5 Q$ can be calculated by the same procedure. Thus in the following calculations we will mainly focus on the extraction of vector form factors $f_i$ while the axial-vector form factors $g_i$ can be extracted analogously. 

At QCD level, the correlation function is calculated by OPE and the use of Eq.~(\ref{quarksmatrix}), the  $J^{V}_{\mu}(x)$ induced correlation function can be expressed as
\begin{eqnarray}
\Pi_{\mu, V}^{QCD}(p_{\Sigma},q) & =&\frac{i}{4}\int d^{4}x\int_{0}^{\infty}d\omega\omega\int_{0}^{1}due^{i(q+\bar{u}\omega v)\cdot x}\nonumber\\
 && \times\Big\{v_{+}\psi_{\parallel}^{n*}(\omega,u)f^{(1)}\bar{u}_{\Sigma}[\bar{\slashed v}\gamma_{5}\gamma^{\nu}CS^{Q}(x)^{T}C^{T}\gamma_{\mu}\bar{\slashed n}\gamma_{\nu}\gamma_{5}]\nonumber\\
 & &-\psi_{\parallel}^{n\bar{n}*}(\omega,u)f^{(2)}\bar{u}_{\Sigma}[\bar{\slashed v}\gamma_{5}\gamma^{\nu}CS^{Q}(x)^{T}C^{T}\gamma_{\mu}i\sigma^{\alpha\beta}\gamma_{\nu}\gamma_{5}]\bar{n}_{\alpha}n_{\beta}\nonumber\\
 & &-2\psi_{\parallel}^{1*}(\omega,u)f^{(2)}\bar{u}_{\Sigma}[\bar{\slashed v}\gamma_{5}\gamma^{\nu}CS^{Q}(x)^{T}C^{T}\gamma_{\mu}\gamma_{\nu}\gamma_{5}]\nonumber\\
 & &-\frac{1}{v_{+}}\psi_{\parallel}^{\bar{n}*}(\omega,u)f^{(1)}\bar{u}_{\Sigma}[\bar{\slashed v}\gamma_{5}\gamma^{\nu}CS^{Q}(x)^{T}C^{T}\gamma_{\mu}\slashed n\gamma_{\nu}\gamma_{5}]\Big\},\label{Corre1}
\end{eqnarray}
where $S^{Q}(x)$ is the QCD free heavy quark propagator.
$\bar v$  and the light-cone vectors $n$, $\bar n$ in Eqs.~(\ref{Corre1}) should be expressed in Lorentz covariant forms
\begin{equation}
n_{\mu} =\frac{1}{v\cdot x}x_{\mu},\ \ \ \bar{n}_{\mu}=2v_{\mu}-\frac{1}{v\cdot x}x_{\mu},\ \ \ \bar{v}_{\mu}=\frac{x_{\mu}}{v\cdot x}-v_{\mu}.\label{nnbar}
\end{equation}
The correlation function can be written as a convolution of the LCDAs and the perturbation kernel, where diquark momenta $\omega$ and momenta fraction $u$ are integrated
\begin{eqnarray}
\Pi_{\mu,V}^{QCD}(p_{\Sigma},q) & =&\frac{1}{4}\int d^{4}x\int_{0}^{2s_0}d\omega\omega\int_{0}^{1}du\int\frac{d^4k}{(2\pi)^4}e^{i(q+\bar{u}\omega v-k)\cdot x}\nonumber\\
 && \times\Big\{\psi_{2}(\omega,u)f^{(1)}\bar{u}_{\Sigma_{c}}\big[\big(\frac{\slashed  x}{v \cdot x}-\slashed v\big) \gamma_{5} \gamma^{\nu} \frac{\slashed  k-m_{Q}}{k^{2}-m_{Q}^{2}} \gamma_{\mu}\big(2 \slashed  v-\frac{\slashed x}{v \cdot x}\big) \gamma_{\nu} \gamma_{5}\big]\nonumber\\
 && -\psi_{3\sigma}(\omega,u)f^{(2)}\bar{u}_{\Sigma_{c}}\big[\big(\frac{\slashed x}{v \cdot x}-\slashed v\big) \gamma_{5} \gamma^{\nu} \frac{\slashed k-m_{Q}}{k^{2}-m_{Q}^{2}} \gamma_{\mu} i \sigma^{\alpha \beta}\big(2 v_{\alpha}-\frac{x_{\alpha}}{v \cdot x}\big) \frac{x_{\beta}}{v \cdot x} \gamma_{\nu} \gamma_{5}\big]\nonumber\\
 && -2\psi_{3s}(\omega,u)f^{(2)}\bar{u}_{\Sigma_{c}}\big[\big(\frac{\slashed{x}}{v \cdot x}-\slashed v\big) \gamma_{5} \gamma^{\nu} \frac{\slashed k-m_{Q}}{k^{2}-m_{Q}^{2}} \gamma_{\mu} \gamma_{\nu} \gamma_{5}\big]\nonumber\\
 && +\psi_{4}(\omega,u)f^{(1)}\bar{u}_{\Sigma_{c}}\big[\big(\frac{\slashed{x}}{v \cdot x}-\slashed v\big) \gamma_{5} \gamma^{\nu} \frac{\slashed k-m_{Q}}{k^{2}-m_{Q}^{2}} \gamma_{\mu} \frac{\slashed x}{v\cdot x} \gamma_{\nu}\gamma_{5}\big]\Big\}.
\end{eqnarray}
 After integrating out the space-time coordinate $x$, one arrive at the explicit form of the correlation function at QCD level
\begin{eqnarray}
&&\Pi_{\mu,V}^{QCD}((p_{\Sigma}+q)^2,q^2)\nonumber\\
 &=&\int_{0}^{2s_0} d \omega \int_{0}^{1} d u f^{(1)}\left\{-\frac{1}{2} \bar{u}^{2} \hat{\psi}_{ \|}^{n *}(\omega, u) \left[\frac{8\bar{u}_{\Sigma}N_{1}}{[(q+\bar{u} \omega v)^{2}-m_{Q}^{2}]^{3}} -\frac{2\bar{u}_{\Sigma}N_{2}}{[(q+\bar{u} \omega v)^{2}-m_{Q}^{2}]^{2}} \right]\right.\nonumber\\
& &+\left.\frac{1}{2}\bar{u} \tilde{\psi}_{ \|}^{n *}(\omega, u)\left[-\frac{2\bar{u}_{\Sigma}N_{3}}{[(q+\bar{u} \omega v)^{2}-m_{Q}^{2}]^{2}} -\frac{2\bar{u}_{\Sigma}\gamma_{\mu}}{(q+\bar{u} \omega v)^{2}-m_{Q}^{2}} \right]
 + \omega \psi_{ \|}^{n *}(\omega, u) \frac{\bar{u}_{\Sigma}N_4}{(q+\bar{u} \omega v)^{2}-m_{Q}^{2}} \right\}\nonumber\\
 &+&\int_{0}^{2s_0} d \omega \int_{0}^{1} d u f^{(2)}\left\{\frac{1}{2}  \bar{u}^{2} \hat{\psi}_{ \|}^{n \bar{n} *}(\omega, u) \left[\frac{8\bar{u}_{\Sigma}N_5}{[(q+\bar{u} \omega v)^{2}-m_{Q}^{2}]^{3}} -\frac{2\bar{u}_{\Sigma}N_6}{[(q+\bar{u} \omega v)^{2}-m_{Q}^{2}]^{2}}\right]\right.\nonumber\\
 &&\left.+\bar{u} \tilde{\psi}_{ \|}^{n \bar{n}}(\omega, u) \left[-\frac{\bar{u}_{\Sigma}[\slashed q,\slashed v](2q_{\mu}+2\bar{u}\omega v_{\mu}-\gamma_{\mu}m_{Q})}{[(q+\bar{u} \omega v)^{2}-m_{Q}^{2}]^{2}} +\frac{\bar{u}_{\Sigma}[\gamma_{\mu},\slashed v]}{(q+\bar{u} \omega v)^{2}-m_{Q}^{2}}\right] \right\}\nonumber\\
 &+&\int_{0}^{2s_0} d \omega \int_{0}^{1} d u f^{(2)} \left\{\bar{u}\tilde{\psi}_{1}^{1 *}(\omega, u)\left[ -\frac{2\bar{u}_{\Sigma}N_7}{[(q+\bar{u} \omega v)^{2}-m_{Q}^{2}]^{2}} +\frac{2\bar{u}_{\Sigma}\gamma_{\mu}}{(q+\bar{u} \omega v)^{2}-m_{Q}^{2}} \right]\right.\nonumber\\
 &&\left.+ \omega\psi_{ \|}^{1 *}(\omega, u)\frac{\bar{u}_{\Sigma}\left(2 q_{\mu}+2 \bar{u} \omega v_{\mu}-\gamma_{\mu} m_{Q}\right)}{(q+\bar{u} \omega v)^{2}-m_{Q}^{2}} \right\}\nonumber\\
 &+&\int_{0}^{2s_0} d \omega \int_{0}^{1} d u f^{(1)}\left\{-\frac{1}{2} \bar{u}^{2} \hat{\psi}_{ \|}^{\bar{n} *}(\omega, u) \left[\frac{8\bar{u}_{\Sigma}N_{8}}{[(q+\bar{u} \omega v)^{2}-m_{Q}^{2}]^{3}} +\frac{4\bar{u}_{\Sigma}N_{9}}{[(q+\bar{u} \omega v)^{2}-m_{Q}^{2}]^{2}} \right]\right.\nonumber\\
&&\left.+\frac{1}{2} \bar{u} \tilde{\psi}_{ \|}^{\bar{n} *}(\omega, u) \left[\frac{2\bar{u}_{\Sigma}N_{10}}{[(q+\bar{u} \omega v)^{2}-m_{Q}^{2}]^{2}} +\frac{2\bar{u}_{\Sigma}\gamma_{\mu}}{(q+\bar{u} \omega v)^{2}-m_{Q}^{2}}  \right]\right\}
 ,\label{correQCD}
\end{eqnarray}
where
\begin{eqnarray}
N_1&=&(q+\bar{u} \omega v)^{2} \gamma_{\mu}(\slashed q+\bar{u} \omega \slashed v)-2m_Q (q_{\mu}+\bar u \omega v_{\mu})(\slashed q+\bar{u} \omega \slashed v),\nonumber\\
N_2&=&6 \gamma_{\mu}(\slashed q+\bar{u} \omega \slashed v)-2 m_{Q} \gamma_{\mu},\nonumber\\
N_3&=&(\slashed q+\bar{u} \omega \slashed v)\gamma_{\mu}(\slashed q+\bar{u} \omega \slashed v)+4 m_{Q} v_{\mu}(\slashed q+\bar{u} \omega \slashed v)+2 m_{Q}\left(q_{\mu}+\bar{u} \omega v_{\mu}\right),\nonumber\\
N_4&=& \gamma_{\mu} (\slashed q+\bar u\omega\slashed v)+2 m_{Q} v_{\mu},\nonumber\\
N_5&=&(\slashed{q}+\bar{u} \omega \slashed v)[\slashed q, \slashed v]\left(2 q_{\mu}+2 \bar{u} \omega v_{\mu}-\gamma_{\mu} m_{Q}\right),\nonumber\\
N_6&=&\gamma_{\rho}\left[\gamma^{\rho}, \slashed v\right]\left(2 q_{\mu}+2 \bar{u} \omega v_{\mu}-\gamma_{\mu} m_{Q}+2 \gamma_{\mu}[\slashed q, \slashed v]\right)+2(\slashed q+\bar{u} \omega\slashed v)\left[\gamma_{\mu}, \slashed v\right],\nonumber\\
N_7&=&(\slashed q+\bar{u} \omega \slashed v)\left(2 q_{\mu}+2 \bar{u} \omega v_{\mu}-\gamma_{\mu} m_{Q}\right),\nonumber\\
N_8&=&(q+\bar{u} \omega v)^{2} \gamma_{\mu}(\slashed q+\bar{u} \omega \slashed v)-2 m_{Q}\left(q_{\mu}+\bar{u} \omega v_{\mu}\right)(\slashed q+\bar{u} \omega \slashed v),\nonumber\\
N_9&=&(\slashed q+\bar{u} \omega \slashed v) \gamma_{\mu}-2 \gamma_{\mu}(\slashed q+\bar{u} \omega \slashed v)-2\left(q_{\mu}+\bar{u} \omega v_{\mu}\right)+m_{Q} \gamma_{\mu},\nonumber\\
N_{10}&=&(\slashed q+\bar{u} \omega \slashed v) \gamma_{\mu}(\slashed q+\bar{u} \omega \slashed v)-2 m_{Q}\left(q_{\mu}+\bar{u} \omega v_{\mu}\right),
\end{eqnarray}
$m_Q$ is the mass of the translating heavy quark, and
\begin{eqnarray}
&&[(q+\bar{u} \omega v)^{2}-m_{Q}^{2}]^{2}=\frac{\bar{u}\omega}{m_{\Sigma}}s
+H(u,\omega,q^{2})-m_{Q}^2,\nonumber\\
&&s=(p_{\Sigma}+q)^2,\ \ 
H(u,\omega,q^{2})=\bar{u}\omega(\bar{u}\omega-m_{\Sigma})+(1-\frac{\bar{u}\omega}{m_{\Sigma}})q^{2}.
\end{eqnarray}

Here we have also used two kinds of newly defined LCDAs
\begin{eqnarray}
\tilde{\psi}_i(\omega,u) =\int_{0}^{\omega}d\tau\tau\psi_i(\tau,u),\ \ \ 
 \hat{\psi}_i(\omega, u) =\int_{0}^{\omega} d \tau \tilde{\psi}_i(\tau, u) \ \ \ \ (i=2,\ 3\sigma,\ 3s,\ 4).\nonumber
\end{eqnarray}

\begin{figure}
\includegraphics[width=0.4\columnwidth]{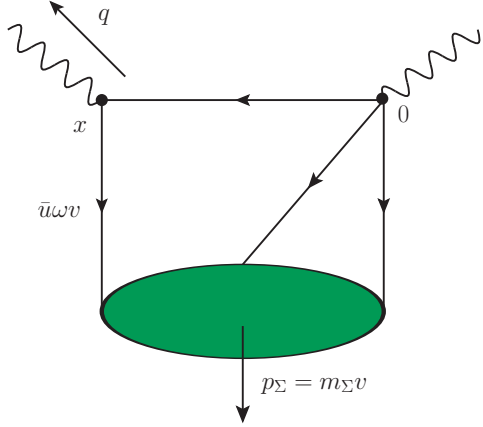} 
\caption{Feynman diagram of the QCD level correlation function. The green ellipse denotes the final $\Sigma_{Q^{\prime}}$ baryon with four-velocity $v$. The left black dot denotes the $V-A$ current while the right dot denotes the doubly-heavy baryon current. The left straight line denote one of the light quark inside the $\Sigma_{Q^{\prime}}$. It has momentum $\bar u \omega v$, where $\bar u$ is its momentum fraction of the diquark momentum. }
\label{fig:FeymDiag} 
\end{figure}
As a result, the correlation function at QCD level can be expressed by six Dirac structures $\{\gamma_{\mu}, v_{\mu}, q_{\mu}, \gamma_{\mu}\slashed q, v_{\mu}\slashed q, q_{\mu}\slashed q \}$. Denoting the corresponding coefficients as $C_{\rm structure}$, one have
\begin{eqnarray}
&&\Pi_{\mu,V}^{QCD}((p_{\Sigma}+q)^2,q^2)\nonumber\\
 &=&\bar{u}_{\Sigma}(p_{\Sigma})\{C_{\gamma_{\mu}}\gamma_{\mu}+C_{v_{\mu}}v_{\mu}+C_{q_{\mu}}q_{\mu}+C_{\gamma_{\mu}\slashed q}\gamma_{\mu}\slashed q+C_{v_{\mu}\slashed q}v_{\mu}\slashed q+C_{q_{\mu}\slashed q}q_{\mu}\slashed q\}.
 \end{eqnarray}
With the denotion $(q+\bar{u} \omega v)^{2}-m_{Q}^{2}=\Delta$, the expression of the coefficients are
\begin{eqnarray}
C_{\gamma_{\mu}}&=&\int_{0}^{2s_0} d \omega \int_{0}^{1} d u\Big\{
\big[{f^{(1)}} {\bar{u}} ({\tilde{\psi}^{\bar{n}*}_{\parallel}}-{\psi^{n*}_{\parallel}} \omega 
  ^2-{\tilde{\psi}^{n*}_{\parallel}})-{f^{(2)}} ({m_{Q}} {\psi^{1*}_{\parallel}}  \omega +2 
  {\tilde{\psi}^{n\bar{n}*}_{\parallel}} {\bar{u}}-2 {\tilde{\psi}^{1*}_{\parallel}} {\bar{u}})
\big]\frac{1}{\Delta}\nonumber\\
&&+{\bar{u}} \big[-2 {f^{(1)}} {\bar{u}} ({\hat{\psi}^{\bar{n}*}_{\parallel}}+{\hat{\psi}^{n*}_{\parallel}}) 
  ({m_{Q}}+3 {\bar{u}} \omega )+{f^{(1)}} ({\tilde{\psi}^{n*}_{\parallel}}-{\tilde{\psi}^{\bar{n}*}_{\parallel}}) 
  \left(q^2+{\bar{u}} \omega  (2 {q\cdot v}+{\bar{u}} \omega 
  )\right)\nonumber\\
  &&\quad+2 {f^{(2)}} {m_{Q}} (3 
  {\hat{\psi}^{n\bar{n}*}_{\parallel}} {\bar{u}}+{\tilde{\psi}^{n\bar{n}*}_{\parallel}} {q\cdot v}
  +{\tilde{\psi}^{1*}_{\parallel}} 
  {\bar{u}} \omega )+4 {f^{(2)}} {\hat{\psi}^{n\bar{n}*}_{\parallel}} {\bar{u}} 
  ({q\cdot v}+{\bar{u}} \omega )\big]
\frac{1}{\Delta^2}\nonumber\\
&&+
4 {\bar{u}}^2 \big[{f^{(1)}} {\bar{u}} \omega 
  ({\hat{\psi}^{\bar{n}*}_{\parallel}}+{\hat{\psi}^{n*}_{\parallel}}) \left(q^2+{\bar{u}} \omega  (2 
  {q\cdot v}+{\bar{u}} \omega )\right)-2 {f^{(2)}} {m_{Q}} 
  {\hat{\psi}^{n\bar{n}*}_{\parallel}} \left(q^2+{q\cdot v} {\bar{u}} \omega \right)\big]
{\frac{1}{\Delta^3}}\Big\},\nonumber\\
\label{eq:cgamu} \\
C_{q_{\mu}}&=&\int_{0}^{2s_0} d \omega \int_{0}^{1} d u{\frac{2 {f^{(2)}} {\psi^{1*}_{\parallel}} \omega}{\Delta}}
+\big[-2 {\bar{u}} ({f^{(1)}} ({m_{Q}} ({\tilde{\psi}^{\bar{n}*}_{\parallel}}+{\tilde{\psi}^{n*}_{\parallel}})+{\bar{u}} 
  \omega  ({\tilde{\psi}^{n*}_{\parallel}}-{\tilde{\psi}^{\bar{n}*}_{\parallel}}))\nonumber\\
  &&+2 {f^{(2)}} ({m_{Q}} 
  ({\tilde{\psi}^{n\bar{n}*}_{\parallel}}-{\tilde{\psi}^{1*}_{\parallel}})+5 {\hat{\psi}^{n\bar{n}*}_{\parallel}} 
  {\bar{u}}+{\tilde{\psi}^{n\bar{n}*}_{\parallel}} {q\cdot v}+{\tilde{\psi}^{1*}_{\parallel}} {\bar{u}} \omega ))
\big]{\frac{1}{\Delta^2}}\nonumber\\
&&+8 {\bar{u}}^2 \big[{f^{(1)}} {m_{Q}} {\bar{u}} \omega  
  ({\hat{\psi}^{\bar{n}*}_{\parallel}}+{\hat{\psi}^{n*}_{\parallel}})+2 {f^{(2)}} {\hat{\psi}^{n\bar{n}*}_{\parallel}} 
  \left({\bar{u}} \omega  ({m_{Q}}+{q\cdot v})+{m_{Q}} 
  {q\cdot v}+q^2\right)\big]
{\frac{1}{\Delta^3}},\label{eq:cqmu}
\\
C_{v_{\mu}}&=&\int_{0}^{2s_0} d \omega \int_{0}^{1} d u\Big\{
2 \big[{f^{(1)}} {\psi^{n*}_{\parallel}} \omega  ({m_{Q}}+{\bar{u}} \omega )+{f^{(2)}} 
  {\bar{u}} \left({\tilde{\psi}^{n\bar{n}*}_{\parallel}}+{\psi^{1*}_{\parallel}} \omega ^2\right)\big]
{\frac{1}{\Delta}}\nonumber\\
&&-2 {\bar{u}}^2 \omega  \big\{{f^{(1)}} \big[{m_{Q}} ({\tilde{\psi}^{\bar{n}*}_{\parallel}}+3 
  {\tilde{\psi}^{n*}_{\parallel}})-6 {\bar{u}} ({\hat{\psi}^{\bar{n}*}_{\parallel}}+{\hat{\psi}^{n*}_{\parallel}})+{\bar{u}} 
  \omega  ({\tilde{\psi}^{n*}_{\parallel}}-{\tilde{\psi}^{\bar{n}*}_{\parallel}})\big]\nonumber\\
  &&\quad+2 {f^{(2)}} (4 {\hat{\psi}^{n\bar{n}*}_{\parallel}} 
  {\bar{u}}+{\tilde{\psi}^{n\bar{n}*}_{\parallel}} {q\cdot v}+{\tilde{\psi}^{1*}_{\parallel}} {\bar{u}} \omega )
\big\}{\frac{1}{\Delta^2}}\nonumber\\
&&-8 {\bar{u}}^3 \omega  \big[{f^{(1)}} ({\hat{\psi}^{\bar{n}*}_{\parallel}}+{\hat{\psi}^{n*}_{\parallel}}) 
  \left({\bar{u}} \omega  (-{m_{Q}}+2 {q\cdot v}+{\bar{u}} \omega 
  )+q^2\right)-2 {f^{(2)}} {\hat{\psi}^{n\bar{n}*}_{\parallel}} \left(q^2+{q\cdot v} {\bar{u}} 
  \omega \right)\big]
{\frac{1}{\Delta^3}}\Big\},\nonumber\\
\label{eq:cvmu}
\\
C_{\gamma_{\mu}q\!\!\!\slash}&=&\int_{0}^{2s_0} d \omega \int_{0}^{1} d u\Big\{
{\frac{{f^{(1)}} {\psi^{n*}_{\parallel}} \omega}{\Delta}}+2 {\bar{u}} [3 {f^{(1)}} {\bar{u}} ({\hat{\psi}^{\bar{n}*}_{\parallel}}+{\hat{\psi}^{n*}_{\parallel}})+{f^{(2)}} 
  {m_{Q}} ({\tilde{\psi}^{n\bar{n}*}_{\parallel}}-{\tilde{\psi}^{1*}_{\parallel}})]
{\frac{1}{\Delta^2}}
\nonumber\\
&&-4 {\bar{u}}^2 \Big[{f^{(1)}} ({\hat{\psi}^{\bar{n}*}_{\parallel}}+{\hat{\psi}^{n*}_{\parallel}}) 
  \left(q^2+{\bar{u}} \omega  (2 {q\cdot v}+{\bar{u}} \omega )\right)+2 
  {f^{(2)}} {m_{Q}} {\hat{\psi}^{n\bar{n}*}_{\parallel}} ({q\cdot v}+{\bar{u}} \omega )\Big]
{\frac{1}{\Delta^3}}\Big\},\label{eq:cgamuqslash}
\\
C_{q_{\mu}q\!\!\!\slash}&=&\int_{0}^{2s_0} d \omega \int_{0}^{1} d u\Big\{
2 {\bar{u}} [{f^{(1)}} ({\tilde{\psi}^{\bar{n}*}_{\parallel}}-{\tilde{\psi}^{n*}_{\parallel}})+2 {f^{(2)}} 
  ({\tilde{\psi}^{n\bar{n}*}_{\parallel}}-{\tilde{\psi}^{1*}_{\parallel}})]
{\frac{1}{\Delta^2}}\nonumber\\
&&+8 {\bar{u}}^2 [{f^{(1)}} {m_{Q}} ({\hat{\psi}^{\bar{n}*}_{\parallel}}+{\hat{\psi}^{n*}_{\parallel}})-2 
  {f^{(2)}} {\hat{\psi}^{n\bar{n}*}_{\parallel}} ({q\cdot v}+{\bar{u}} \omega )]
{\frac{1}{\Delta^3}}\Big\},\label{eq:cqmuqslash}
\\
C_{v_{\mu}q\!\!\!\slash}&=&\int_{0}^{2s_0} d \omega \int_{0}^{1} d u\Big\{
2 {\bar{u}} \big\{{f^{(1)}} [{\bar{u}} \omega  ({\tilde{\psi}^{\bar{n}*}_{\parallel}}-{\tilde{\psi}^{n*}_{\parallel}})-2 
  {m_{Q}} {\tilde{\psi}^{n*}_{\parallel}}]+2 {f^{(2)}} {\bar{u}} [{\hat{\psi}^{n\bar{n}*}_{\parallel}}+\omega  
  ({\tilde{\psi}^{n\bar{n}*}_{\parallel}}-{\tilde{\psi}^{1*}_{\parallel}})]\big\}
{\frac{1}{\Delta^2}}\nonumber\\
&&+8 {\bar{u}}^3 \omega  [{f^{(1)}} {m_{Q}} ({\hat{\psi}^{\bar{n}*}_{\parallel}}+{\hat{\psi}^{n*}_{\parallel}})-2 
  {f^{(2)}} {\hat{\psi}^{n\bar{n}*}_{\parallel}} ({q\cdot v}+{\bar{u}} \omega )
]{\frac{1}{\Delta^3}}\Big\},\label{eq:cvmuqslash}
\end{eqnarray}
where we have omitted the arguments $(\omega,u)$ of the LCDAs $\psi_{\parallel}$, $\tilde{\psi}_{\parallel}$ and $\hat{\psi}_{\parallel}$ for compactness.
Fig.~\ref{fig:FeymDiag} shows the Feynman diagram describing the QCD level correlation function. Since the correlation function is a function of Lorentz invariants $(p_{\Sigma}+q)^2$ and $q^2$. By extracting the discontinuity of the correlation function Eq.~(\ref{correQCD}) acrossing the branch cut on the $(p_{\Sigma}+q)^2$ complex plane, the correlation function can be expressed as a dispersion integration form
\begin{eqnarray}
\Pi_{\mu,V}^{QCD}(p_{\Sigma},q)=\frac{1}{\pi}\int_{(m_{Q}+m_{Q^{\prime}}+m_{q})^{2}}^{\infty}ds\frac{\rm{Im}\Pi_{\mu,V}^{QCD}(s,q^2)}{s-(p_{\Sigma}+q)^{2}}.\label{DiscorreQCD}
\end{eqnarray}
According to the global Quark-Hadron duality, the continuum contribution
Eq.~(\ref{correHadron}) should be equal to the QCD level contribution Eq.~(\ref{DiscorreQCD}) in the same spectral region $s_{th}<s<\infty$. 

The equivalence between Eq. (\ref{DiscorreQCD}) and Eq. (\ref{Formextract}) enables one to extract the form factors $F_i^{+}$
\begin{eqnarray}
& &-\frac{f_{\Xi}}{(q+p_{\Sigma})^{2}-m_{\Xi}^{2}}\bar{u}_{\Sigma}(p_{\Sigma})F_{1}^{+}(q^{2}) =  \frac{1}{\pi}\int_{(m_{Q}+m_{Q^{\prime}}+m_{q})^{2}}^{s_{th}}\frac{ds}{s-(p_{\Sigma}+q)^{2}}\rm{Im}\Big\{\frac{C_{\gamma_{\mu}}+C_{\gamma_{\mu}\slashed q} (m_{\Sigma}+m_{\Xi})}{2 m_{\Xi}}\Big\},\nonumber\\
&&\label{fromfactor1} \\
& &-\frac{f_{\Xi}}{(q+p_{\Sigma})^{2}-m_{\Xi}^{2}}\bar{u}_{\Sigma}(p_{\Sigma})F_{2}^{+}(q^{2})  = \frac{1}{\pi}\int_{(m_{Q}+m_{Q^{\prime}}+m_{q})^{2}}^{s_{th}}\frac{ds}{s-(p_{\Sigma}+q)^{2}}\nonumber\\
&&\times\frac{1}{2 m_{\Sigma}m_{\Xi}}\rm{Im}\left[-2 m_{\Sigma}C_{\gamma_{\mu}\slashed q} +(m_{\Sigma}^2-m_{\Sigma}m_{\Xi}) C_{q_{\mu}\slashed q}+(m_{\Xi}-m_{\Sigma})C_{v_{\mu}\slashed q}-m_{\Sigma} C_{q_{\mu}}+C_{v_{\mu}}\right],
\label{fromfactor2}\\
& &-\frac{f_{\Xi}}{(q+p_{\Sigma})^{2}-m_{\Xi}^{2}}\bar{u}_{\Sigma}(p_{\Sigma})F_{3}^{+}(q^{2}) =  \frac{1}{\pi}\int_{(m_{Q}+m_{Q^{\prime}}+m_{q})^{2}}^{s_{th}}\frac{ds}{s-(p_{\Sigma}+q)^{2}}\rm{Im}\Big\{\frac{(m_{\Xi}-m_{\Sigma}) C_{\gamma_{\mu}\slashed q} +C_{q_{\mu}}}{2 m_{\Xi}}\Big\},\nonumber\\
&&\label{fromfactor3} 
\end{eqnarray}
which describing the positive parity baryon $\Xi^{P+}_{QQ^{\prime}}$ decays. By constructing Borel transformation on the both sides of the three Eqs.~(\ref{fromfactor1})-(\ref{fromfactor3}), one can get the explicit expression of each form factors $F_i^{+}$. Before Borel transformation, we make the following transform on the denominators in Eqs.~(\ref{eq:cgamu})-(\ref{eq:cvmuqslash}) to enable the extraction of the discontinuity,
\begin{eqnarray}
{\frac{1}{[(q+\bar{u}\omega v)^2-m_{Q}^2]^n}}\to\frac{1}{(n-1)!}\left(\frac{\partial}{\partial \Omega}\right)^{(n-1)}{\frac{1}{(q+\bar{u}\omega v)^2-\Omega}}\Big|_{\Omega=m_Q^2}.
\end{eqnarray}
Due to the complex form of the analytic expression,
here we just show the analytic expression of $F_{1}^{+}$ as an example
\begin{eqnarray}
& &F_{1}^{+}(q^{2}) = -\frac{1}{f_{\Xi}}exp(\frac{m_{\Xi}^2}{M^2}) 
\int_{0}^{2s_0} d \omega \int_{0}^{1} d u\frac{m_{\Sigma}}{\bar{u}\omega}exp(-\frac{s_{r}}{M^2})
\theta(sth-s_{r})\theta(s_{r}-(m_{Q}+m_{Q^{\prime}}+m_{q})^{2})\nonumber\\
&&\times\Big\{\big[{f^{(1)}} {\bar{u}} ({\tilde{\psi}^{\bar{n}*}_{\parallel}}-{\psi^{n*}_{\parallel}} \omega 
  ^2-{\tilde{\psi}^{n*}_{\parallel}})-{f^{(2)}} ({m_{Q}} {\psi^{1*}_{\parallel}}  \omega +2 
  {\tilde{\psi}^{n\bar{n}*}_{\parallel}} {\bar{u}}-2 {\tilde{\psi}^{1*}_{\parallel}} {\bar{u}})
\big]\frac{1}{2m_{\Xi}}+f^{(1)} \psi^{n*}_{\parallel}\frac{m_{\Sigma}+m_{\Xi}}{2m_{\Xi}}\Big\}\nonumber\\
&&-\frac{1}{f_{\Xi}}exp(\frac{m_{\Xi}^2}{M^2}) \frac{\partial}{\partial \Omega}
\int_{0}^{2s_0} d \omega \int_{0}^{1} d u\frac{m_{\Sigma}}{\bar{u}\omega}exp(-\frac{s_{r}^{\Omega}}{M^2})
\theta(sth-s_{r}^{\Omega})\theta(s_{r}^{\Omega}-(\sqrt{\Omega}+m_{Q^{\prime}}+m_{q})^{2})\nonumber\\
&&\times\Big\{{\bar{u}} \big[-2 {f^{(1)}} {\bar{u}} ({\hat{\psi}^{\bar{n}*}_{\parallel}}+{\hat{\psi}^{n*}_{\parallel}}) 
  ({m_{Q}}+3 {\bar{u}} \omega )+{f^{(1)}} ({\tilde{\psi}^{n*}_{\parallel}}-{\tilde{\psi}^{\bar{n}*}_{\parallel}}) 
  \left(q^2+{\bar{u}} \omega  (2 {q\cdot v}+{\bar{u}} \omega 
  )\right)\nonumber\\
  &&\quad+2 {f^{(2)}} {m_{Q}} (3 
  {\hat{\psi}^{n\bar{n}*}_{\parallel}} {\bar{u}}+{\tilde{\psi}^{n\bar{n}*}_{\parallel}} {q\cdot v}
  +{\tilde{\psi}^{1*}_{\parallel}} 
  {\bar{u}} \omega )+4 {f^{(2)}} {\hat{\psi}^{n\bar{n}*}_{\parallel}} {\bar{u}} 
  ({q\cdot v}+{\bar{u}} \omega )\big]\frac{1}{2m_{\Xi}}\nonumber\\
  &&\quad+2 {\bar{u}} [3 {f^{(1)}} {\bar{u}} ({\hat{\psi}^{\bar{n}*}_{\parallel}}+{\hat{\psi}^{n*}_{\parallel}})+{f^{(2)}} 
  {m_{Q}} ({\tilde{\psi}^{n\bar{n}*}_{\parallel}}-{\tilde{\psi}^{1*}_{\parallel}})]\frac{m_{\Sigma}+m_{\Xi}}{2m_{\Xi}}\Big\}\Big|_{\Omega=m_Q^2}\nonumber\\
  &&-\frac{1}{f_{\Xi}}exp(\frac{m_{\Xi}^2}{M^2}) \frac{1}{2}\left(\frac{\partial}{\partial \Omega}\right)^2
\int_{0}^{2s_0} d \omega \int_{0}^{1} d u\frac{m_{\Sigma}}{\bar{u}\omega}exp(-\frac{s_{r}^{\Omega}}{M^2})
\theta(sth-s_{r}^{\Omega})\theta(s_{r}^{\Omega}-(m_{Q}+m_{Q^{\prime}}+m_{q})^{2})\nonumber\\
&&\times\Big\{4 {\bar{u}}^2 \left[{f^{(1)}} {\bar{u}} \omega 
  ({\hat{\psi}^{\bar{n}*}_{\parallel}}+{\hat{\psi}^{n*}_{\parallel}}) \left(q^2+{\bar{u}} \omega  (2 
  {q\cdot v}+{\bar{u}} \omega )\right)-2 {f^{(2)}} {m_{Q}} 
  {\hat{\psi}^{n\bar{n}*}_{\parallel}} \left(q^2+{q\cdot v} {\bar{u}} \omega \right)\right]\frac{1}{2m_{\Xi}}\nonumber\\
&&\quad-4 {\bar{u}}^2 \left[{f^{(1)}} ({\hat{\psi}^{\bar{n}*}_{\parallel}}+{\hat{\psi}^{n*}_{\parallel}}) 
  \left(q^2+{\bar{u}} \omega  (2 {q\cdot v}+{\bar{u}} \omega )\right)+2 
  {f^{(2)}} {m_{Q}} {\hat{\psi}^{n\bar{n}*}_{\parallel}} ({q\cdot v}+{\bar{u}} \omega )\right] \frac{m_{\Sigma}+m_{\Xi}}{2m_{\Xi}}\Big\}\Big|_{\Omega=m_Q^2},\nonumber\\
   \label{fromfactor1} 
  \end{eqnarray}
where $s_r$ and $s^{\Omega}_r$ are defined as the singularity position of the heavy quark propagator in the correlation function. In other words, they are the roots of the following two equations respectively
\begin{eqnarray}
&&\frac{\bar{u}\omega}{m_{\Sigma}}s_r+H(u,\omega,q^{2})-m_{Q}^2=0,\nonumber\\
&&\frac{\bar{u}\omega}{m_{\Sigma}}s^{\Omega}_r+H(u,\omega,q^{2})-\Omega=0.
\end{eqnarray}
Further, since the $G_i^{+}$ can be obtained in a similar way, we will not show them explicitly here.

\section{Numerical results}
\label{sec:numerical results}
\subsection{Transition Form Factors}

In this work, the heavy quark masses are taken as $m_{c}=(1.35\pm0.10)\ {\rm GeV}$
and $m_{b}=(4.7\pm0.1)\ {\rm GeV}$ while the masses of light quarks are neglected. The masses, lifetimes and decay constants $f_{\Xi}$ of doubly heavy baryons are shown in Tab.~\ref{Tab:para_if}~\cite{Karliner:2014gca,Shah:2016vmd,Shah:2017liu,Kiselev:2001fw}. While the masses and decay constants of $\Sigma_Q$ are taken as $m_{\Sigma_c}=2.454$ GeV, $m_{\Sigma_b}=5.814$ GeV, and $f^{(1)}=f^{(2)}=0.038$~\cite{Groote:1997yr}. The upper limit of light quarks momentum is  $s_0=1.2$ GeV~\cite{Ali:2012pn}.

\begin{table}[!htb]
\caption{Masses, lifetimes and decay constants of doubly heavy baryons.}
\label{Tab:para_if} %
\begin{tabular}{c|c|c|c}
\hline 
\hline
Baryons & Mass (GeV) & Lifetime (fs) &$f_{\Xi}$ $({\rm GeV}^{3})$~\cite{Hu:2017dzi}\tabularnewline
\hline  
$\Xi_{cc}^{++}$  & $3.621$ \cite{Aaij:2017ueg}  & 256 ~\cite{Aaij:2018wzf}  & $0.109\pm0.021$ \tabularnewline
$\Xi_{cc}^{+}$  & $3.621$ \cite{Aaij:2017ueg}  & 45~\cite{Cheng:2018mwu} & $0.109\pm0.021$ \tabularnewline
$\Xi_{bc}^{+}$  & $6.943$ \cite{Brown:2014ena}  & 244  \cite{Karliner:2014gca}& $0.176\pm0.040$ \tabularnewline
$\Xi_{bc}^{0}$  & $6.943$ \cite{Brown:2014ena}  & 93 \cite{Karliner:2014gca}& $0.176\pm0.040$ \tabularnewline
$\Xi_{bb}^{0}$  & $10.143$ \cite{Brown:2014ena}  & 370 \cite{Karliner:2014gca}& $0.281\pm0.071$ \tabularnewline
$\Xi_{bb}^{-}$  & $10.143$ \cite{Brown:2014ena}  & 370 \cite{Karliner:2014gca}& $0.281\pm0.071$ \tabularnewline
\hline 
\hline
\end{tabular}
\end{table}
\begin{table}
\caption{Threshold $s_{th}$ of $\Xi_{QQ^{\prime}}$, Borel parameters $M^2$, and $q^2$ range for fitting form factors.}
\label{Tab:LCSRpars}
\begin{center}
\begin{tabular}{c|c|c|c}
\hline 
\hline 
Channel & $s_{th}$ (GeV$^{2}$) & $M^{2}$ (GeV$^{2}$) & Fit Range (GeV$^{2}$)\tabularnewline
\hline 
$\Xi_{cc}\to\Sigma_{c}$ & $16\pm1$ & $15\pm1$ & $0<q^{2}<0.8$\tabularnewline
$\Xi_{bb}\to\Sigma_{b}$ & $112\pm2$ & $20\pm1$ & $0<q^{2}<6$\tabularnewline
$\Xi_{bc}\to\Sigma_{c}$ & $54\pm1.5$ & $20\pm1$ & $0<q^{2}<6$\tabularnewline
$\Xi_{bc}\to\Sigma_{b}$ & $54\pm1.5$ & $20\pm1$ & $0<q^{2}<0.8$\tabularnewline
\hline 
\hline 
\end{tabular}
\end{center}
\end{table}

The threshold $s_{th}$ of $\Xi_{QQ^{\prime}}$ and Borel parameters $M^2$ are shown in Tab.~\ref{Tab:LCSRpars}, which are chosen as to make the form factors stable.  Since the light-cone OPE for heavy baryon transition is reliable in the region where $q^2$ is positive but not too large, to parameterize the form factors one needs to limit the $q^2$ region which are listed in the last column of Tab.~\ref{Tab:LCSRpars}. The parameterization formula used in this work is
\begin{equation}
	F(q^{2})=\frac{F(0)[1+a (q^2)+b (q^2)^2]}{1-\frac{q^{2}}{m_{{\rm fit}}^{2}}+\delta\left(\frac{q^{2}}{m_{{\rm fit}}^{2}}\right)^{2}}\label{eq:fit_formula_1},
\end{equation}
where the denominator reflects the pole structure of the form factors, while the nominator is due to the assumption that the form factors have a polynomial form at small $q^2$.
The numerical and fitting results for the form factors are given in Tab.~\ref{Tab:ff_cc_bb_bc}, where the \textcolor{red}{``}Null" means that the corresponding parameter is set to be zero before the fitting.

\begin{table}
\caption{Form factors of the transition $\Xi_{QQ^{\prime}q_{2}}\to\Sigma_{Q^{\prime}q_{1}q_{2}}$.
$F(0)$, $m_{fit}$, $\delta$, $a$ and $b$ correspond to the five
fitting parameters in Eq.~(\ref{eq:fit_formula_1}). The form factors
of $\Xi_{QQ^{\prime}q}\to\Sigma_{Q^{\prime}qq}$ are just $\sqrt{2}$
times those of $\Xi_{QQ^{\prime}q_{2}}\to\Sigma_{Q^{\prime}q_{1}q_{2}}$,
which are not shown explicitly in this table.}
\label{Tab:ff_cc_bb_bc} %
\begin{tabular}{c|c|c|c|c|c}
\hline 
\hline 
$F$  & $F(0)$  & $m_{{\rm {fit}}}$  & $\delta$  & $a$  & $b$\tabularnewline
\hline 
$f_{1}^{\Xi_{cc}^{++}\to\Sigma_{c}^{+}}$  & $-0.86\pm0.08$  & $1.22\pm0.06$  & $0.46\pm0.02$  & Null  & Null \tabularnewline
$f_{2}^{\Xi_{cc}^{++}\to\Sigma_{c}^{+}}$  & $-1.19\pm0.06$  & $1.20\pm0.01$  & $0.25\pm0.01$  & $-0.67\pm0.02$ & $-0.18\pm0.0$\tabularnewline
$f_{3}^{\Xi_{cc}^{++}\to\Sigma_{c}^{+}}$  & $-0.66\pm0.17$  & $0.93\pm0.06$  & $0.43\pm0.03$  & Null  & Null \tabularnewline
$g_{1}^{\Xi_{cc}^{++}\to\Sigma_{c}^{+}}$  & $0.07\pm0.02$  & $1.24\pm0.01$  & $0.21\pm0.01$  & $1.15\pm0.28$  & $1.88\pm0.36$\tabularnewline
$g_{2}^{\Xi_{cc}^{++}\to\Sigma_{c}^{+}}$  & $2.08\pm0.1$  & $0.85\pm0.12$  & $0.26\pm0.04$  & $-1.63\pm0.47$  & $0.66\pm0.39$ \tabularnewline
$g_{3}^{\Xi_{cc}^{++}\to\Sigma_{c}^{+}}$  & $7.0\pm0.02$  & $1.13\pm0.03$  & $0.24\pm0.0$  & Null  & Null \tabularnewline
\hline 
$f_{1}^{\Xi_{bc}^{+}\to\Sigma_{b}^{0}}$  & $-0.38\pm0.03$  & $1.14\pm0.02$  & $0.48\pm0.01$  & Null  & Null \tabularnewline
$f_{2}^{\Xi_{bc}^{+}\to\Sigma_{b}^{0}}$  & $-1.46\pm0.06$  & $2.32\pm1.15$  & $-6.37\pm6.71$  & $-0.18\pm0.55$ & $-0.86\pm0.60$\tabularnewline
$f_{3}^{\Xi_{bc}^{+}\to\Sigma_{b}^{0}}$  & $-0.84\pm0.08$  & $0.89\pm0.01$  & $0.46\pm0.01$  & Null  & Null \tabularnewline
$g_{1}^{\Xi_{bc}^{+}\to\Sigma_{b}^{0}}$  & $0.05\pm0.01$  & $1.6\pm0.54$  & $0.89\pm0.36$  & $1.92\pm0.45$ & $-2.54\pm0.77$\tabularnewline
$g_{2}^{\Xi_{bc}^{+}\to\Sigma_{b}^{0}}$  & $4.46\pm0.22$  & $1.98\pm0.01$  & $1.17\pm0.22$  & $-0.59\pm0.00$  & $0.07\pm0.02$ \tabularnewline
$g_{3}^{\Xi_{bc}^{+}\to\Sigma_{b}^{0}}$  & $11.64\pm0.35$  & $0.94\pm0.02$  & $0.28\pm0.02$  & Null  & Null \tabularnewline
\hline 
$f_{1}^{\Xi_{bc}^{0}\to\Sigma_{c}^{+}}$  & $-0.21\pm0.01$  & $6.48\pm0.0$  & $-0.41\pm0.01$  & Null  & Null \tabularnewline
$f_{2}^{\Xi_{bc}^{0}\to\Sigma_{c}^{+}}$  & $-0.09\pm0.0$  & $3.43\pm6.86$  & $0.45\pm0.0$  & Null  & Null \tabularnewline
$f_{3}^{\Xi_{bc}^{0}\to\Sigma_{c}^{+}}$  & $0.03\pm0.0$  & $3.3\pm0.04$  & $0.43\pm0.02$  & $-0.33\pm0.02$  & $0.01\pm0.0$\tabularnewline
$g_{1}^{\Xi_{bc}^{0}\to\Sigma_{c}^{+}}$  & $0.01\pm0.0$  & $3.96\pm0.93$  & $0.42\pm0.22$  & $1.58\pm0.53$  & $-0.08\pm0.02$\tabularnewline
$g_{2}^{\Xi_{bc}^{0}\to\Sigma_{c}^{+}}$  & $0.12\pm0.01$  & $4.43\pm0.21$  & $1.01\pm0.32$  & $-0.06\pm0.0$  & $0.0\pm0.0$ \tabularnewline
$g_{3}^{\Xi_{bc}^{0}\to\Sigma_{c}^{+}}$  & $0.47\pm0.02$  & $2.78\pm0.01$  & $0.3\pm0.0$  & $-0.12\pm0.0$ & $0.01\pm0.0$ \tabularnewline
\hline 
$f_{1}^{\Xi_{bb}^{-}\to\Sigma_{b}^{0}}$  & $-0.3\pm0.01$  & $3.5\pm0.15$  & $1.5\pm0.19$  & $-0.07\pm0.01$ & $0.01\pm0.0$\tabularnewline
$f_{2}^{\Xi_{bb}^{-}\to\Sigma_{b}^{0}}$  & $0.1\pm0.01$  & $5.6\pm1.72$  & $2.85\pm2.02$  & $-0.28\pm0.03$ & $0.0\pm0.01$\tabularnewline
$f_{3}^{\Xi_{bb}^{-}\to\Sigma_{b}^{0}}$  & $-0.26\pm0.01$  & $3.56\pm0.01$  & $0.52\pm0.05$  & $0.07\pm0.01$ & $-0.01\pm0.0$\tabularnewline
$g_{1}^{\Xi_{bb}^{-}\to\Sigma_{b}^{0}}$  & $-0.01\pm0.01$  & $4.55\pm3.87$  & $1.4\pm1.43$  & $-2.0\pm6.6$ & $0.12\pm12.27$\tabularnewline
$g_{2}^{\Xi_{bb}^{-}\to\Sigma_{b}^{0}}$  & $0.05\pm0.02$  & $4.35\pm0.11$  & $1.21\pm0.07$  & $-0.32\pm0.03$ & $0.07\pm0.01$\tabularnewline
$g_{3}^{\Xi_{bb}^{-}\to\Sigma_{b}^{0}}$  & $1.02\pm0.13$  & $3.26\pm0.12$  & $0.56\pm0.07$  & $-0.32\pm0.01$ & $0.02\pm0.0$\tabularnewline
\hline 
\hline 
\end{tabular}
\end{table}

Since we are interested in the error coming from our theoretical approach, the errors of the form factors are estimated from the thresholds $s_{th}$ and Borel parameters $M^2$, which are the only tunable parameters in LCSR. The $q^2$ dependence of the form factors corresponding to the four channels are shown in Fig.~\ref{fig:Formfactors}, with the parameters $s_{th}$, $M^2$ fixed at their center values as shown in Tab.~\ref{Tab:LCSRpars}.

 Tabs.~\ref{Tab:comparison_cc} and \ref{Tab:comparison_bb_bc} show the comparison between this work and other works in the previous literatures. From the comparison one can find that the vector form factors $f_i$ obtained in this work fit well with  those of other works, especially that from QCDSR. However, there are certain difference for other form factors $g_{i}$. Some comments on such difference between various works are given as follows.
 
\begin{table}
\caption{Comparison of $\Xi_{cc}$ decay form factors derived in this work
with the results from QCD sum rules (QCDSR)~\cite{Shi:2019hbf},
light-front quark model (LFQM)~\cite{Wang:2017mqp}, the non-relativistic
quark model (NRQM) and the MIT bag model (MBM)~\cite{PerezMarcial:1989yh}.}
\label{Tab:comparison_cc} \centering{}%
\begin{tabular}{c|c|c|c|c|c|c}
\hline 
\hline 
Transitions  & $F(0)$  & This work  & QCDSR~\cite{Shi:2019hbf}  & LFQM~\cite{Wang:2017mqp}  & NRQM ~\cite{PerezMarcial:1989yh}  & MBM ~\cite{PerezMarcial:1989yh} \tabularnewline
\hline 
$\Xi_{cc}^{++}\to\Sigma_{c}^{+}$  & $f_{1}(0)$  & $-0.86\pm0.08$  & $-0.35\pm0.04$  & $-0.46$  & $-0.28$  & $-0.30$\tabularnewline
 & $f_{2}(0)$  & $-1.19\pm0.06$  & $1.15\pm0.12$  & $1.04$  & $0.14$  & $0.91$\tabularnewline
 & $f_{3}(0)$  & $-0.66\pm0.17$  & $-1.40\pm0.39$  & - -  & $-0.10$  & $0.07$\tabularnewline
 & $g_{1}(0)$  & $0.07\pm0.02$  & $-0.23\pm0.06$  & $-0.62$  & $-0.70$  & $-0.56$\tabularnewline
 & $g_{2}(0)$  & $2.08\pm0.1$  & $-0.26\pm0.15$  & $0.04$  & $-0.02$  & $0.05$\tabularnewline
 & $g_{3}(0)$  & $7.0\pm0.02$  & $2.68\pm0.39$  & - -  & $0.10$  & $2.59$\tabularnewline
\hline 
\hline 
\end{tabular}
\end{table}

\begin{table}
\caption{Comparison of $\Xi_{bb}$ and $\Xi_{bc}$ decay form factors derived
in this work with the results from QCD sum rules (QCDSR)~\cite{Shi:2019hbf}
and light-front quark model (LFQM)~\cite{Wang:2017mqp}.}
\label{Tab:comparison_bb_bc} \centering{}%
\begin{tabular}{c|c|c|c|c}
\hline 
\hline 
Transitions  & $F(0)$  & This work  & QCDSR~\cite{Shi:2019hbf}  & LFQM~\cite{Wang:2017mqp} \tabularnewline
\hline 
$\Xi_{bc}^{+}\to\Sigma_{b}^{0}$  & $f_{1}(0)$  & $-0.38\pm0.03$  & $-0.28\pm0.03$  & $-0.32$ \tabularnewline
 & $f_{2}(0)$  & $-1.46\pm0.06$  & $2.04\pm0.21$  & $1.54$ \tabularnewline
 & $f_{3}(0)$  & $-0.84\pm0.08$  & $-3.78\pm1.38$  & - - \tabularnewline
 & $g_{1}(0)$  & $0.05\pm0.01$  & $-0.13\pm0.06$  & $-0.41$ \tabularnewline
 & $g_{2}(0)$  & $4.46\pm0.22$  & $-0.18\pm0.25$  & $0.18$ \tabularnewline
 & $g_{3}(0)$  & $11.64\pm0.35$  & $10.1\pm1.4$  & - - \tabularnewline
\hline 
$\Xi_{bc}^{0}\to\Sigma_{c}^{+}$  & $f_{1}(0)$  & $-0.21\pm0.01$  & $-0.22\pm0.03$  & $-0.07$ \tabularnewline
 & $f_{2}(0)$  & $-0.09\pm0.0$  & $0.36\pm0.06$  & $0.10$ \tabularnewline
 & $f_{3}(0)$  & $0.03\pm0.0$  & $-0.45\pm0.07$  & - - \tabularnewline
 & $g_{1}(0)$  & $0.01\pm0.0$  & $-0.22\pm0.03$  & $-0.10$ \tabularnewline
 & $g_{2}(0)$  & $0.12\pm0.01$  & $-0.31\pm0.05$  & $-0.003$ \tabularnewline
 & $g_{3}(0)$  & $0.47\pm0.02$  & $0.47\pm0.07$  & - - \tabularnewline
\hline 
$\Xi_{bb}^{-}\to\Sigma_{b}^{0}$  & $f_{1}(0)$  & $-0.3\pm0.01$  & $-0.12\pm0.01$  & $-0.06$ \tabularnewline
 & $f_{2}(0)$  & $0.1\pm0.01$  & $0.22\pm0.03$  & $0.15$ \tabularnewline
 & $f_{3}(0)$  & $-0.26\pm0.01$  & $-0.46\pm0.06$  & - - \tabularnewline
 & $g_{1}(0)$  & $-0.01\pm0.01$  & $-0.12\pm0.01$  & $-0.09$ \tabularnewline
 & $g_{2}(0)$  & $0.05\pm0.02$  & $-0.19\pm0.03$  & $-0.02$ \tabularnewline
 & $g_{3}(0)$  & $1.02\pm0.13$  & $0.49\pm0.07$  & - - \tabularnewline
\hline 
\hline 
\end{tabular}
\end{table}
 
In terms of this work and the work based on QCDSR, the difference between the form factors derived from the two approaches can be attributed to following two points: 
\begin{itemize}

\item  In the QCDSR work, the authors performed a leading order calculation for a three-point correlation function by OPE, where all the non-perturbative effects are produced by the condensates of the dimension 3 to 5 operators. In the LCSR work, the non-perturbative effects are produced by the LCDAs of $\Sigma_Q$ baryons, which were also derived by QCDSR~\cite{Ali:2012pn}, but with only dimension 3 condensates included. As a result, the amount of non-perturbative effects introduced in this work and the QCDSR work are different. 

\item Ideally, when all order QCD corrections and the complete series of OPE are included, the results from QCDSR and LCSR calculation should be equivalent. However, both in the QCDSR and LCSR works, only leading order calculations are performed. In addition, the QCDSR work only contains contribution from several low dimensional operator condensates, while in this work, only several leading twist LCDAs are introduced. Thus we have in fact extracted two parts of the same form factor respectively in the two works. Generally, these two parts will overlap but will not be the same. 

\end{itemize}

In terms of this work and the work based on LFQM, the difference between the form factors derived from the two approaches can be attributed to two points: 
\begin{itemize}

\item In this work, the hadron transition matrix element is extracted from a correlation function by finding the mass pole residues, where uncertainties may occur when one \textcolor{red}{separates} the single particle state from the continuum spectrum. However, the LFQM directly expresses the initial and final hadronic states in terms of the quark level wave functions. This operation avoids the potential pollution from heavier spectrum but introduces more tunable parameters, which may be inconsistent with LCSR.

\item The baryon pictures between LCSR and LFQM are very different. In LCSR, the $\Sigma_Q$ baryons are described by their LCDAs, which are defined in full QCD theory and describe a three-body system. In LFQM, the baryons are described by a two-body Gaussian shape wave function, where the two spectator quarks are combined to be a point-like diquark. 

\end{itemize}
 
\begin{figure}
\includegraphics[width=1.0\columnwidth]{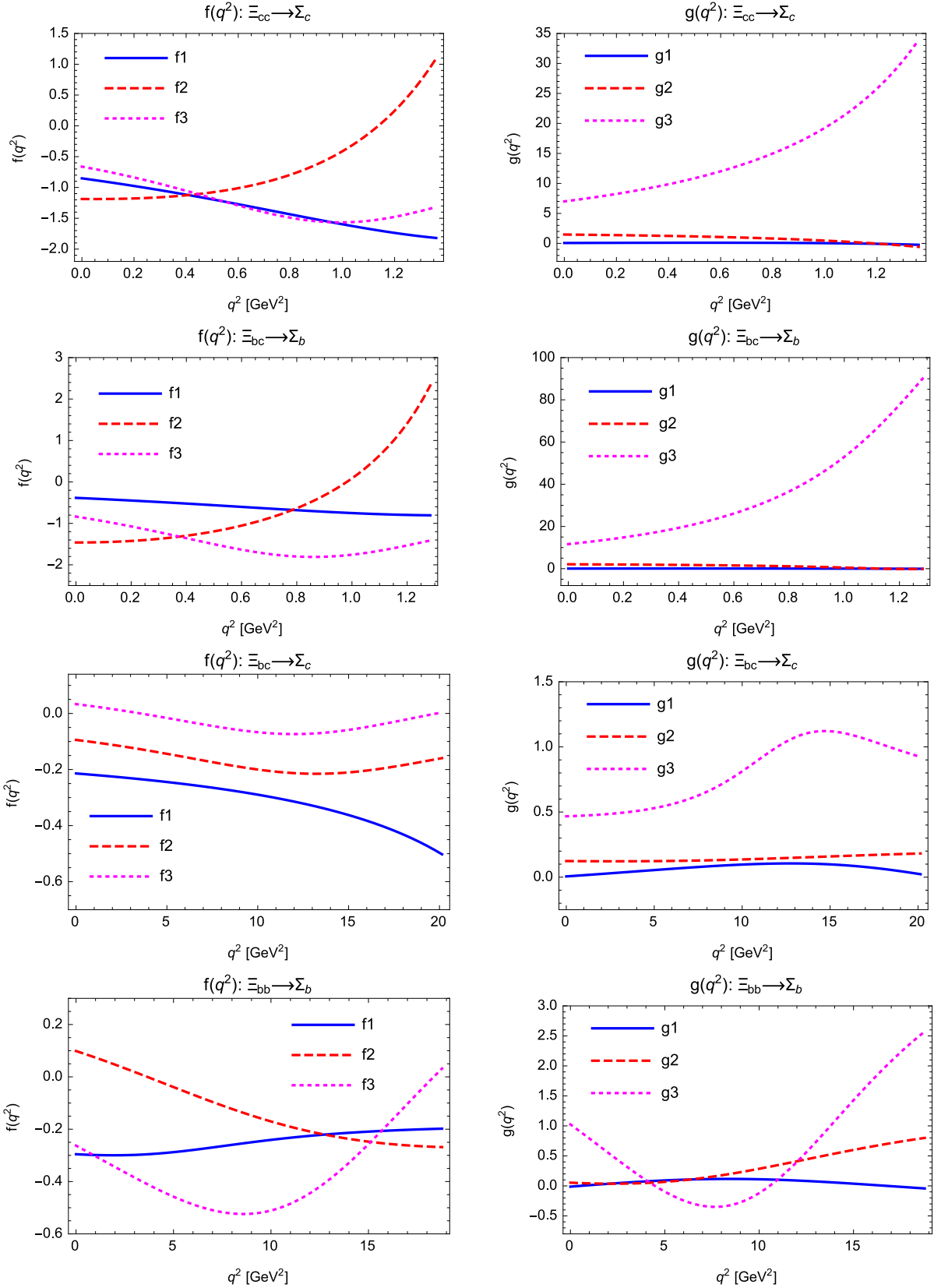} 
\caption{$q^2$ dependence of the $\Xi_{QQ^{\prime}q_{2}}\to \Sigma_{Q^{\prime}q_{1}q_{2}}$ form factors. The first two graphs correspond to $\Xi_{cc}\to\Sigma_c$, the second two graphs correspond to $\Xi_{bc}\to\Sigma_b$, the third two graphs correspond to $\Xi_{bc}\to\Sigma_c$ and the fourth two graphs correspond to $\Xi_{bb}\to\Sigma_b$. Here the parameters $s_{th}$, $M^2$ are fixed at their center values as shown in Tab.~\ref{Tab:LCSRpars}. For the case of $\Xi_{QQ^{\prime}q}\to\Sigma_{Q^{\prime}qq}$, the vertical scale needs to be enlarged by a factor $\sqrt{2}$.}
\label{fig:Formfactors} 
\end{figure}
		

\subsection{Semi-leptonic Decays}
The effective Hamiltonian inducing the semi-leptonic decays $\Xi_{QQ^{\prime}}\to\Sigma_{Q^{\prime}}$ is
\begin{eqnarray}
{\cal H}_{{\rm eff}} & = & \frac{G_{F}}{\sqrt{2}}\bigg(V_{ub}[\bar{u}\gamma_{\mu}(1-\gamma_{5})b][\bar{l}\gamma^{\mu}(1-\gamma_{5})\nu]+V_{cd}^{*}[\bar{d}\gamma_{\mu}(1-\gamma_{5})c][\bar{\nu}\gamma^{\mu}(1-\gamma_{5})l]\bigg),
\end{eqnarray}
where the Fermi constant $G_{F}$ and CKM matrix elements are taken from Refs.~\cite{Olive:2016xmw,Tanabashi:2018oca}
\begin{align}
&G_{F}=1.166\times10^{-5}{\rm GeV}^{-2},\quad
|V_{ub}|=0.00357,\quad|V_{cd}|=0.225.\label{eq:GFCKM}
\end{align}

The decay amplitudes induced by vector current and axial-vector current can be expressed in terms of the  following helicity amplitudes
\begin{eqnarray}
H_{\frac{1}{2},0}^{V} & = & -i\frac{\sqrt{Q_{-}}}{\sqrt{q^{2}}}\left((M_{1}+M_{2})f_{1}-\frac{q^{2}}{M_{1}}f_{2}\right),\;\;\;
H_{\frac{1}{2},0}^{A} =  -i\frac{\sqrt{Q_{+}}}{\sqrt{q^{2}}}\left((M_{1}-M_{2})g_{1}+\frac{q^{2}}{M}g_{2}\right),\nonumber \\
H_{\frac{1}{2},1}^{V} & = & i\sqrt{2Q_{-}}\left(-f_{1}+\frac{M_{1}+M_{2}}{M_{1}}f_{2}\right),\;\;\;
H_{\frac{1}{2},1}^{A}  =  i\sqrt{2Q_{+}}\left(-g_{1}-\frac{M_{1}-M_{2}}{M_{1}}g_{2}\right),\nonumber \\
H_{\frac{1}{2},t}^{V} & = & -i\frac{\sqrt{Q_{+}}}{\sqrt{q^{2}}}\left((M_{1}-M_{2})f_{1}+\frac{q^{2}}{M_{1}}f_{3}\right),\;\;\;
H_{\frac{1}{2},t}^{A} =  -i\frac{\sqrt{Q_{-}}}{\sqrt{q^{2}}}\left((M_{1}+M_{2})g_{1}-\frac{q^{2}}{M_{1}}g_{3}\right),
\end{eqnarray}
where $Q_{\pm}=(M_1\pm M_2)^{2}-q^{2}$. $M_{1}$ and $M_{2}$ are the masses of the initial and final baryon. The amplitudes with negative helicity external states have the following simple relations with those having positive helicities
\begin{equation}
H_{-\lambda_{2},-\lambda_{W}}^{V}=H_{\lambda_{2},\lambda_{W}}^{V}\quad\text{and}\quad H_{-\lambda_{2},-\lambda_{W}}^{A}=-H_{\lambda_{2},\lambda_{W}}^{A},
\end{equation}
$\lambda_{2}$ and $\lambda_{W}$ denotes the polarizations of the final $\Sigma_{Q^{\prime}}$ and the intermediate $W$ boson, respectively. The total helicity amplitudes induced by the $V-A$ current are  
\begin{equation}
H_{\lambda_{2},\lambda_{W}}=H_{\lambda_{2},\lambda_{W}}^{V}-H_{\lambda_{2},\lambda_{W}}^{A}.
\end{equation}

\begin{figure}
\includegraphics[width=1.0\columnwidth]{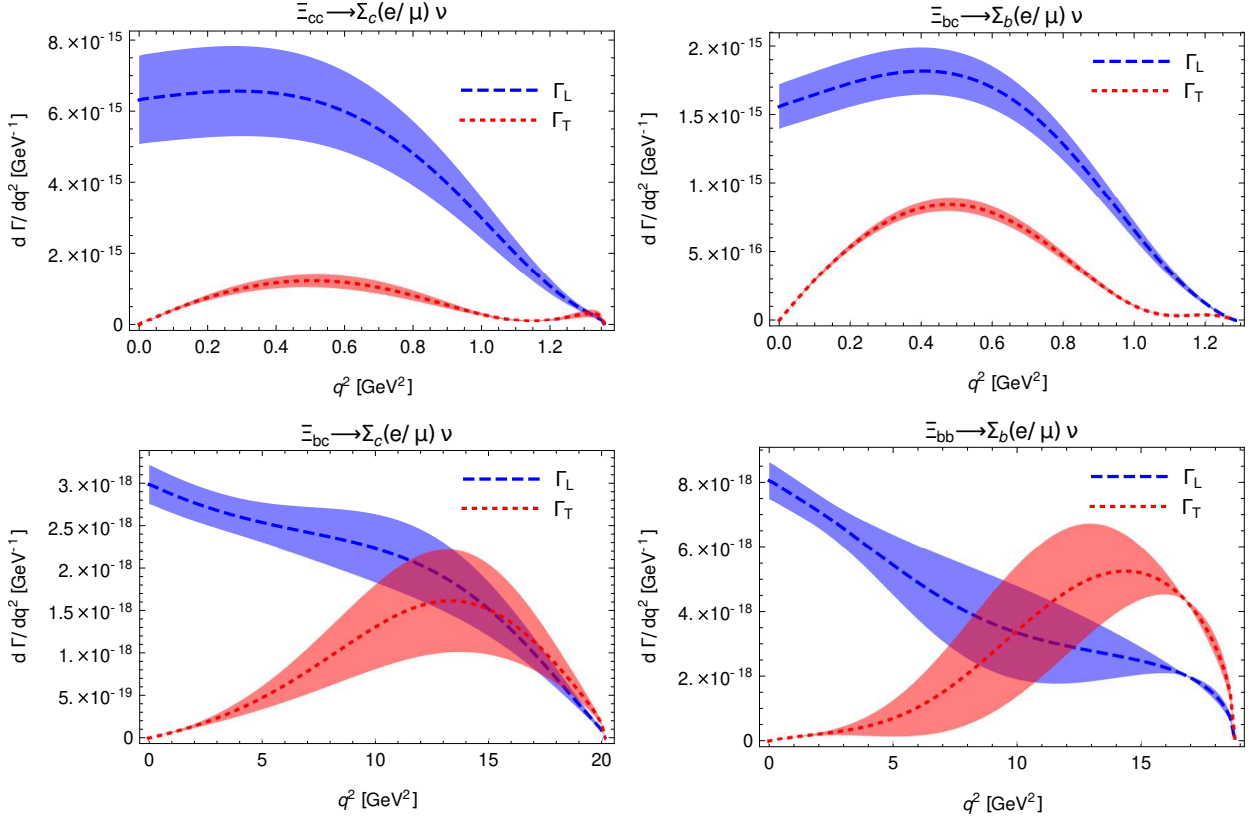} 
\caption{$q^2$ dependence of the semi-leptonic $\Xi_{QQ^{\prime}q_2}\to\Sigma_{Q^{\prime}q_1q_2}l\nu_l$ decay widths. $d\Gamma_L/dq^2$ are shown by the blue bands while $d\Gamma_T/dq^2$ are shown by the red bands. The dashed lines are the center value curves and the band width reflects the corresponding error. For the case of $\Xi_{QQ^{\prime}q}\to\Sigma_{Q^{\prime}qq}$, the vertical scale needs to be enlarged by a factor $2$.}
\label{fig:SemiDecayWidths} 
\end{figure}

Decay widths of $\Xi_{QQ^{\prime}}\to\Sigma_{Q^{\prime}}l\nu$ can be classified by the polarization of the $l\nu$ pairs. The \textcolor{blue}{decay widths} with longitudinally or transversely polarized $l\nu$ pairs are
\begin{align}
\frac{d\Gamma_{L}}{dq^{2}} & =\frac{G_{F}^{2}|V_{{\rm CKM}}|^{2}q^{2}\ p\ (1-\hat{m}_{l}^{2})^{2}}{384\pi^{3}M_{1}^{2}}\left((2+\hat{m}_{l}^{2})(|H_{-\frac{1}{2},0}|^{2}+|H_{\frac{1}{2},0}|^{2})+3\hat{m}_{l}^{2}(|H_{-\frac{1}{2},t}|^{2}+|H_{\frac{1}{2},t}|^{2})\right),\label{eq:longi-1}\\
\frac{d\Gamma_{T}}{dq^{2}} & =\frac{G_{F}^{2}|V_{{\rm CKM}}|^{2}q^{2}\ p\ (1-\hat{m}_{l}^{2})^{2}(2+\hat{m}_{l}^{2})}{384\pi^{3}M_{1}^{2}}(|H_{\frac{1}{2},1}|^{2}+|H_{-\frac{1}{2},-1}|^{2}),\label{eq:trans-1}
\end{align}
where $\hat{m}_{l}\equiv m_{l}/\sqrt{q^{2}}$. Note that when calculating form factors with LCSR, we choose the rest frame of the final baryon $\Sigma_{Q^{\prime}}$. While for simplicity, we calculate the decay width in the rest frame of the initial particle $\Xi_{QQ^{\prime}}$.  $p=\sqrt{Q_{+}Q_{-}}/(2M_{1})$
is the three-momentum magnitude of $\Sigma_{Q^{\prime}}$ in the rest frame of $\Xi_{QQ^{\prime}}$. 
By integrating out the squared transfer momentum $q^{2}$, one can obtain the total decay width 
\begin{equation}
\Gamma=\int_{m_l^2}^{(M_{1}-M_{2})^{2}}dq^{2}\left(\frac{d\Gamma_{L}}{dq^{2}}+\frac{d\Gamma_{T}}{dq^{2}}\right).
\end{equation}
Tab.~\ref{Tab:semi_lep} shows the integrated partial decay widths, branching ratios and the ratios of $\Gamma_{L}/\Gamma_{T}$ for various semi-leptonic $\Xi_{QQ^{\prime}}\to\Sigma_{Q^{\prime}}l\nu_l$ processes, where $l=e,\mu,\tau$. The masses of $e$ and $\mu$ have been neglected while the mass of $\tau$ is taken as $1.78$ GeV~\cite{Olive:2016xmw}. Fig.~\ref{fig:SemiDecayWidths} shows the $q^2$ dependence of the differential decay widths corresponding to four channels. Tab.~\ref{Tab:comparison_semi_lep} gives a comparison of our decay width results with those given in the literatures. It can be found that most of the decay widths from this and other works are of the same order of magnitude.

\begin{table}
\caption{Decay widths and branching ratios of the semi-leptonic $\Xi_{QQ^{\prime}}\to\Sigma_{Q^{\prime}}l\nu_{l}$
decays, where $l=e,\mu$.}
\label{Tab:semi_lep} %
\begin{tabular}{l|c|c|c}
\hline 
\hline 
channels  & $\Gamma/\text{~GeV}$  & ${\cal B}$  & $\Gamma_{L}/\Gamma_{T}$\tabularnewline
\hline 
$\Xi_{cc}^{++}\to\Sigma_{c}^{+}l^{+}\nu_{l}$  & $(7.12\pm1.33)\times10^{-15}$  & $(2.77\pm0.48)\times10^{-3}$  & $6.92\pm2.8$\tabularnewline
$\Xi_{cc}^{+}\to\Sigma_{c}^{0}l^{+}\nu_{l}$  & $(1.42\pm0.26)\times10^{-14}$  & $(9.72\pm1.68)\times10^{-3}$  & $6.92\pm2.8$\tabularnewline
$\Xi_{bc}^{+}\to\Sigma_{b}^{0}l^{+}\nu_{l}$  & $(2.17\pm0.19)\times10^{-15}$  & $(8.06\pm0.7)\times10^{-4}$  & $2.91\pm0.47$\tabularnewline
$\Xi_{bc}^{0}\to\Sigma_{b}^{-}l^{+}\nu_{l}$  & $(4.34\pm0.38)\times10^{-15}$  & $(6.17\pm0.53)\times10^{-4}$  & $2.91\pm0.47$\tabularnewline
$\Xi_{bc}^{0}\to\Sigma_{c}^{+}l^{-}\bar{\nu}_{l}$  & $(5.73\pm1.15)\times10^{-17}$  & $(8.1\pm1.62)\times10^{-6}$  & $2.21\pm1.6$\tabularnewline
$\Xi_{bc}^{0}\to\Sigma_{c}^{+}\tau^{-}\bar{\nu}_{\tau}$  & $(3.5\pm0.62)\times10^{-17}$  & $(4.49\pm0.87)\times10^{-6}$  & $2.45\pm1.68$\tabularnewline
$\Xi_{bc}^{+}\to\Sigma_{c}^{++}l^{-}\bar{\nu}_{l}$  & $(1.14\pm0.23)\times10^{-16}$  & $(4.26\pm0.85)\times10^{-5}$  & $2.21\pm1.6$\tabularnewline
$\Xi_{bc}^{+}\to\Sigma_{c}^{++}\tau^{-}\bar{\nu}_{\tau}$  & $(7.0\pm1.24)\times10^{-17}$  & $(2.6\pm0.46)\times10^{-5}$  & $2.45\pm1.68$\tabularnewline
$\Xi_{bb}^{-}\to\Sigma_{b}^{0}l^{-}\bar{\nu}_{l}$  & $(1.26\pm0.32)\times10^{-16}$  & $(7.05\pm1.82)\times10^{-5}$  & $1.57\pm1.33$\tabularnewline
$\Xi_{bb}^{-}\to\Sigma_{b}^{0}\tau^{-}\bar{\nu}_{\tau}$  & $(6.34\pm1.62)\times10^{-17}$  & $(3.89\pm0.91)\times10^{-5}$  & $1.36\pm1.21$\tabularnewline
$\Xi_{bb}^{0}\to\Sigma_{b}^{+}l^{-}\bar{\nu}_{l}$  & $(2.52\pm0.31)\times10^{-16}$  & $(1.41\pm0.36)\times10^{-4}$  & $1.57\pm1.33$\tabularnewline
$\Xi_{bb}^{0}\to\Sigma_{b}^{+}\tau^{-}\bar{\nu}_{\tau}$  & $(1.26\pm0.32)\times10^{-16}$  & $(7.79\pm1.82)\times10^{-5}$  & $1.36\pm1.21$\tabularnewline
\hline 
\hline 
\end{tabular}
\end{table}

\begin{table}
\caption{Comparison of the decay widths (in units of GeV) for the semi-leptonic
decays in this work with the results derived from QCD sum rules (QCDSR)~\cite{Shi:2019hbf},
the light-front quark model (LFQM)~\cite{Wang:2017mqp}, the heavy
quark spin symmetry (HQSS)~\cite{Albertus:2012nd}, the nonrelativistic
quark model (NRQM)~\cite{PerezMarcial:1989yh} and the MIT bag model
(MBM)~\cite{PerezMarcial:1989yh} in literatures.}
\label{Tab:comparison_semi_lep} \centering{}\resizebox{\textwidth}{28mm}{
\begin{tabular}{c|c|c|c|c|c|c}
\hline 
\hline 
Channels  & This work  & QCDSR~\cite{Shi:2019hbf}  & LFQM~\cite{Wang:2017mqp}  & HQSS~\cite{Albertus:2012nd}  & NRQM~\cite{PerezMarcial:1989yh}  & MBM~\cite{PerezMarcial:1989yh}\tabularnewline
\hline 
$\Xi_{cc}^{++}\to\Sigma_{c}^{+}l^{+}\nu_{l}$  & $(7.12\pm1.33)\times10^{-15}$  & $(2.3\pm0.4)\times10^{-15}$  & $9.60\times10^{-15}$  & $5.22\times10^{-15}$  & $6.58\times10^{-15}$  & $2.63\times10^{-15}$\tabularnewline
\hline 
$\Xi_{cc}^{+}\to\Sigma_{c}^{0}l^{+}\nu_{l}$  & $(1.42\pm0.26)\times10^{-14}$  & $(4.6\pm0.9)\times10^{-15}$  & $1.91\times10^{-14}$  & $1.04\times10^{-14}$  & $1.32\times10^{-14}$  & $5.92\times10^{-15}$\tabularnewline
\hline 
$\Xi_{bb}^{-}\to\Sigma_{b}^{0}l^{-}\bar{\nu}_{l}$  & $(1.26\pm0.32)\times10^{-16}$  & $(1.3\pm0.2)\times10^{-16}$  & $3.33\times10^{-17}$  & - -  & - -  & - -\tabularnewline
\hline 
$\Xi_{bb}^{0}\to\Sigma_{b}^{+}l^{-}\bar{\nu}_{l}$  & $(2.52\pm0.31)\times10^{-16}$  & $(2.5\pm0.4)\times10^{-16}$  & $6.67\times10^{-17}$  & - -  & - -  & - -\tabularnewline
\hline 
$\Xi_{bc}^{0}\to\Sigma_{c}^{+}l^{-}\bar{\nu}_{l}$  & $(5.73\pm1.15)\times10^{-17}$  & $(4.2\pm0.7)\times10^{-16}$  & $4.74\times10^{-17}$  & - -  & - -  & - -\tabularnewline
\hline 
$\Xi_{bc}^{+}\to\Sigma_{c}^{++}l^{-}\bar{\nu}_{l}$  & $(1.14\pm0.23)\times10^{-16}$  & $(8.4\pm1.4)\times10^{-16}$  & $9.48\times10^{-17}$  & - -  & - -  & - -\tabularnewline
\hline 
$\Xi_{bc}^{+}\to\Sigma_{b}^{0}l^{+}\nu_{l}$  & $(2.17\pm0.19)\times10^{-15}$  & $(1.5\pm0.3)\times10^{-15}$  & $4.63\times10^{-15}$  & - -  & - -  & - -\tabularnewline
\hline 
$\Xi_{bc}^{0}\to\Sigma_{b}^{-}l^{+}\nu_{l}$  & $(4.34\pm0.38)\times10^{-15}$  & $(3.0\pm0.5)\times10^{-15}$  & $9.18\times10^{-15}$  & - -  & - -  & - -\tabularnewline
\hline 
\hline 
\end{tabular}} 
\end{table}

For these phenomenology results, here are some remarks:
\begin{itemize}
\item The errors of the decay widths given in Tab.~\ref{Tab:semi_lep} and Fig.~\ref{fig:SemiDecayWidths} totally results from the errors of form factors.

\item From Tab.~\ref{Tab:semi_lep}, one can find that the decay widths of $\Xi_{cc}$ and $\Xi_{bc}\to\Sigma_b$ decays are several orders of magnitude larger than those of $\Xi_{bb}$ decays. 

This phenomenon is mainly due to the huge difference between the CKM matrix elements $|V_{cd}|=0.225 \gg |V_{ub}|=0.00357$. However, the decay width of $\Xi_{bc}\to\Sigma_c$ is also larger than those of $\Xi_{bb}$ decays. This is due to the effect of broken  heavy quark symmetry, which can be easily seen in the view of initial baryons masses
\begin{eqnarray}
&&\frac{\Gamma(\Xi_{bc}\to\Sigma_{c}l\nu_{l})}{\Gamma(\Xi_{bb}\to\Sigma_{b}l\nu_{l})}
\propto(\frac{M_{\Xi_{bb}}}{M_{\Xi_{bc}}})^2>1.\label{eq:bc/bb}
\end{eqnarray}
For the same reason we can understand why $\Gamma(\Xi_{cc}\to\Sigma_{c}l\nu_{l})$ is also larger than $\Gamma(\Xi_{bc}\to\Sigma_{b}l\nu_{l})$. In fact, such feature also emerges in the case of $B$ and $D$ decays.
\item According to the SU(3) symmetry, the decay widths of various semi-leptonic channels are related with each other. Refs.~\cite{Wang:2017azm} and~\cite{Shi:2017dto} have offered a systematic SU(3) analysis of doubly heavy baryon decays as well as a complete decay width relations. Several channels that we have not calculated in this work can still be estimated by SU(3) symmetry~\cite{Wang:2017azm}
\begin{eqnarray}
\Gamma(\Omega_{cc}^{+}\to\Xi_{c}^{\prime0}l^{+}\nu)&=&\Gamma(\Xi_{cc}^{++}\to\Sigma_{c}^{+}l^{+}\nu)=\frac{1}{2}\Gamma(\Xi_{cc}^{+}\to\Sigma_{c}^{0}l^{+}\nu)=(7.12\pm1.33)\times10^{-15} \rm{GeV},\nonumber\\
\Gamma(\Omega_{bc}^{0}\to\Xi_{b}^{\prime-}l^{+}\nu)&=&\Gamma(\Xi_{bc}^{+}\to\Sigma_{b}^{0}l^{+}\nu)=\frac{1}{2}\Gamma(\Xi_{bc}^{0}\to\Sigma_{b}^{-}l^{+}\nu)=(2.17\pm0.19)\times10^{-15} \rm{GeV},\nonumber\\
\Gamma(\Omega_{bc}^{0}\to\Xi_{c}^{\prime+}l^{-}\bar{\nu})&=&\Gamma(\Xi_{bc}^{0}\to\Sigma_{c}^{+}l^{-}\bar{\nu})=\frac{1}{2}\Gamma(\Xi_{bc}^{+}\to\Sigma_{c}^{++}l^{-}\bar{\nu})=(5.73\pm1.15)\times10^{-17} \rm{GeV},\nonumber\\
\Gamma(\Omega_{bb}^{-}\to\Xi_{b}^{\prime0}l^{-}\bar{\nu})&=&\Gamma(\Xi_{bb}^{-}\to\Sigma_{b}^{0}l^{-}\bar{\nu})=\frac{1}{2}\Gamma(\Xi_{bb}^{0}\to\Sigma_{b}^{+}l^{-}\bar{\nu})=(1.26\pm0.32)\times10^{-16} \rm{GeV}.\nonumber\\
\end{eqnarray}

\item From the comparison shown in Tab.~\ref{Tab:comparison_semi_lep}, it seems that our semi-leptonic decay widths fit well with those from other works. Especially, our prediction for $\Gamma(\Xi_{cc}\to\Sigma_{c}l\nu_{l})$ is almost the same as those from LFQM, HQSS, and NRQM calculations. Our prediction for other decay channels fit with either QCDSR or LFQM calculations.

\end{itemize}

\section{Conclusions}
As a continuation of our previous work on $\Xi_{QQ^{\prime}}$ decays into anti-triplets $\Lambda_{Q^{\prime}}$,  this work performs a phenomenological study on the semi-leptonic decay of $\Xi_{QQ^{\prime}}$ into an sextet baryon $\Sigma_{Q^{\prime}}$ with LCSR approach. The transition form factors have been derived with the parallel LCDAs of the final states $\Sigma_{Q^{\prime}}$. With the numerical results of these form factors, the decay widths and branching ratios of the corresponding semi-leptonic process are predicted. The error estimation and theoretical analyses are also given in detail. Our prediction for semi-leptonic decay widths is almost the same as those from most of other works we have compared in this work. Phenomenologically, we hope our predictions for the branching ratios could be tested by LHCb and other experiments in the future. Theoretically, we hope our calculation can help people to exam the singly heavy baryon LCDAs, and further improvement on these LCDAs as well as higher order corrections on this LCSR calculation are also expected in the future.

\label{sec:conclusions}

\section*{Acknowledgements}
The authors are very grateful to Prof. Wei Wang for useful discussions. This work is supported in part by National
Natural Science Foundation of China under Grants No.11575110,  
11735010,  and 11911530088,  Natural Science Foundation of Shanghai under Grants
No.~15DZ2272100, and by Key Laboratory for Particle
Physics, Astrophysics and Cosmology, Ministry of Education.

\end{document}